\documentclass[%
aip,
 amsmath,amssymb,
reprint,%
]{revtex4-1}

\usepackage{graphicx}
\usepackage{dcolumn}
\usepackage{bm}

\usepackage[utf8]{inputenc}
\usepackage[T1]{fontenc}
\usepackage{mathptmx}
\usepackage{etoolbox}
\usepackage{listings}
\usepackage{physics}
\usepackage{booktabs}
\usepackage{tabularx}
\usepackage{ragged2e}  
\usepackage[nolist,nohyperlinks]{acronym}
\usepackage{siunitx}
\usepackage[caption=false]{subfig}
\usepackage{wrapfig}
\usepackage{layouts}
\usepackage[dvipsnames]{xcolor}
\usepackage{hyperref}
\hypersetup{colorlinks=true,allcolors=blue}
\usepackage{floatrow} 
\renewcommand{\thefootnote}{\arabic{footnote}}

\makeatletter
\def\@email#1#2{%
 \endgroup
 \patchcmd{\titleblock@produce}
  {\frontmatter@RRAPformat}
  {\frontmatter@RRAPformat{\produce@RRAP{*#1\href{mailto:#2}{#2}}}\frontmatter@RRAPformat}
  {}{}
}%
\makeatother
\begin{document}


\newcommand{\comment}[1]{\textcolor{ForestGreen}{(#1)}}
\newcommand{\todo}[1]{\textcolor{Red}{(#1)}}

\preprint{AIP/123-QED}

\title[Nonlinear response of telecom-wavelength superconducting single-photon detectors]{Nonlinear response of telecom-wavelength superconducting single-photon detectors}
\author{Patrick Mark}
 \email{patrickmark.pm1@gmail.com}
\author{Sebastian Gstir}
\author{Julian Münzberg}
\author{Gregor Weihs}
\author{Robert Keil}
 \affiliation{Department of Experimental Physics, University of Innsbruck, Technikerstraße 25, 6020 Innsbruck, Austria}
 \email{robert.keil@uibk.ac.at}

\date{\today}

\begin{abstract}
We measure the nonlinearity of a telecom-wavelength superconducting nanowire single-photon detector via incoherent beam combination. At typical photon count rates and detector bias current, the observed relative deviation from a perfectly linear response is in the order of \SI{0.1}{\percent} when the flux is doubled. This arises from a balance between the counteracting nonlinearities of deadtime-induced detector saturation and of multi-photon detections. The observed behaviour is modelled empirically, which suffices for a correction of measured data. In addition, statistical simulations, taking into account the measured recovery of the detection efficiency, provide insight into possible mechanisms of multi-photon detection.

\end{abstract}

\maketitle

%

\begin{acronym}
\acro{snspd}[SNSPD]{superconducting nanowire single-photon detector}
\acro{pd}[PD]{photo diode}
\acro{nl}[NL]{nonlinearity}
\acro{nd}[ND]{neutral density}
\acro{sled}[SLED]{super-luminescent light emitting diode}
\acro{tpd}[TPD]{two-photon detection}
\acro{sde}[SDE]{system detection efficency}
\acro{sem}[SEM]{standard error of mean}
\acro{acem}[ACEM]{autocorrelation corrected error of mean}
\acro{dmm}[DMM]{digital multi-meter}
\acro{fwhm}[FWHM]{full width half maximum}

\end{acronym}


\section{\label{sec:Intro}Introduction}
The efficient detection and counting of single photons is an enabling technology for experiments testing the fundamental principles of quantum mechanics \cite{Christensen2013,Shadbolt2014}, for photonic quantum information processing \cite{Prevedel2007,Kok2016,Flamini2019}, as well as for specific applications such as correlated photon-pair spectroscopy \cite{Whittaker2017,Losero2018}. Superconducting nanowire single-photon detectors (SNSPDs) have revolutionised this process by providing efficient and highly time-resolved detection at a wide range of optical wavelengths \cite{first_snspd,Single-photon_source_and_detectors,Superconducting_Devices_in_Quantum_Optics}. These include the infrared optical telecom bands around \SI{1550}{\nm}, which are preferentially used in fiber-based quantum communication and quantum networks due to their minimal loss at these wavelengths \cite{Maring2017,Chen2020,Wei2022,vanLeent2022}. SNSPDs work via a phase transition between the superconducting and normal conducting state triggered by the absorption of a photon, which can be measured as a voltage pulse across the nanowire.

Linearity is a working assumption of any ideal photodetector, i.e., one assumes that the number of incident photons is proportional to the number of detection events. This is clearly an idealisation as real detectors exhibit nonlinearities. As a consequence, real detectors suffer from a systematic, count-rate dependent deviation from the ideal linear behaviour, which biases all photon-counting measurements. This can be particularly harmful in comparative measurements with countrates varying over a large dynamic range, as arising in quantum state tomography \cite{QST2001,Altepeter2005} or in experimental tests of quantum physics \cite{born2011,Kauten_2017_5Path,Rozema2021,Vogl2021,Gstir_2021,Gstir2023}. Therefore, a precise characterisation of nonlinearities as a function of the incident photon rate is essential to allow for a calibration or compensation of this effect.

For SNSPDs two processes can be expected to produce substantial nonlinear contributions: Dead-time and multi-photon detection. The former arises from the finite recovery time of the supercurrent in the nanowire after a transition \cite{Kerman2006}. This leads to a saturation characteristic, which manifests in a sublinear growth of the detection rate for increasing input photon flux, see Fig.~\ref{fig_sublinear_a}. The corresponding effect in avalanche detectors has been widely investigated \cite{Ware2007,nonlin_pd_Kornilov:14,kautenNonlin,Hlousek2023}. Multi-photon detection, on the other hand, can arise when a single photon fails to trigger a detection, but multiple
arriving photons suffice to cause a phase transition \cite{N_Photon_autocorrelator,P_click}. The likelihood of such multi-photon detection events grows polynomially with the input flux, such that this effect will lead to a supralinear behaviour, see Fig.~\ref{fig_supralinear_b} for an exemplary illustration.

The nonlinearity of detectors is often studied via the superposition method, which operates on the principle of additive intensities in incoherent beam-combination setups \cite{fast_nonlin_det_poly_Coslovi:80}. These studies have allowed for a quantitative analysis of the detector dead-time and the associated saturation in avalanche photodiodes, modelling this effect by a simple step-like detector recovery \cite{kautenNonlin,Hlousek2023}. The superposition method was also used to investigate SNSPDs at near-infrared wavelengths (around \SI{800}{\nm}), revealing intricate dynamics containing both, sub- and supralinear behaviour as a function of photon flux and detector bias current \cite{Hlousek2023}. The origin of these dynamics could, so far, not be determined. 

In this work, we investigate the nonlinearity of an SNSPD at telecom wavelengths via the superposition method. For the default bias current, which sets a good compromise between detection efficiency and dark count rate, we observe a slight supralinear trend for low-to-intermediate count rates (up to $\SI{300}{kHz}$), whereas sublinear saturation dynamics takes over for larger count rates. Via a variation of the bias current, we can attribute these nonlinearities to an interplay of detector dead time and multi-photon detection, in particular two-photon detection. In addition, we go beyond a simple step-function model of the dead time by directly measuring the detector's temporal recovery profile \cite{DirectMeasOfRecoveryTime}. Taking into account this measured recovery profile, we model the multi-photon detections via the temporal dynamics of overlapping resistive hot spots in Monte-Carlo simulations. We also consider the rate-dependent detection efficiency induced by AC-biasing \cite{AC_biasing,Ferrari:19_detect_response_AC_biasing} and show that this effect has only a minor influence in our setting. 

\section{Experimental procedure}\label{Experiment}

\begin{figure}[htb!]
    \sidesubfloat[]{%
     \includegraphics[width=0.35\textwidth]{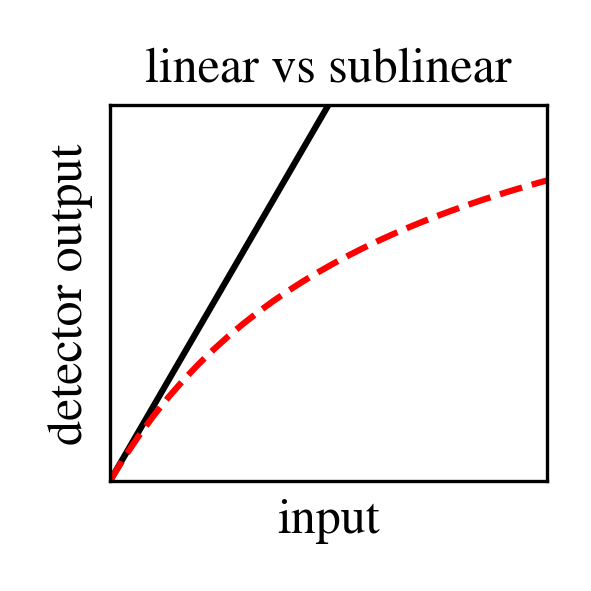}\label{fig_sublinear_a}}
    \sidesubfloat[]{%
    \includegraphics[width=0.35\textwidth]{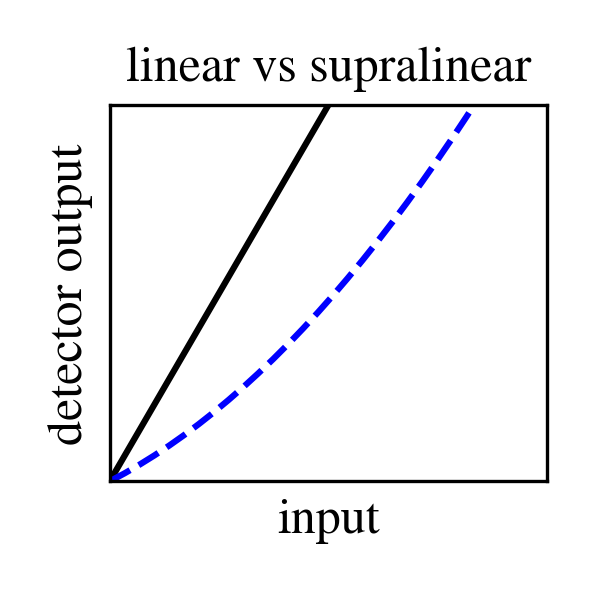}\label{fig_supralinear_b}}
    
    \sidesubfloat[]{%
    \includegraphics[width=0.75\textwidth]{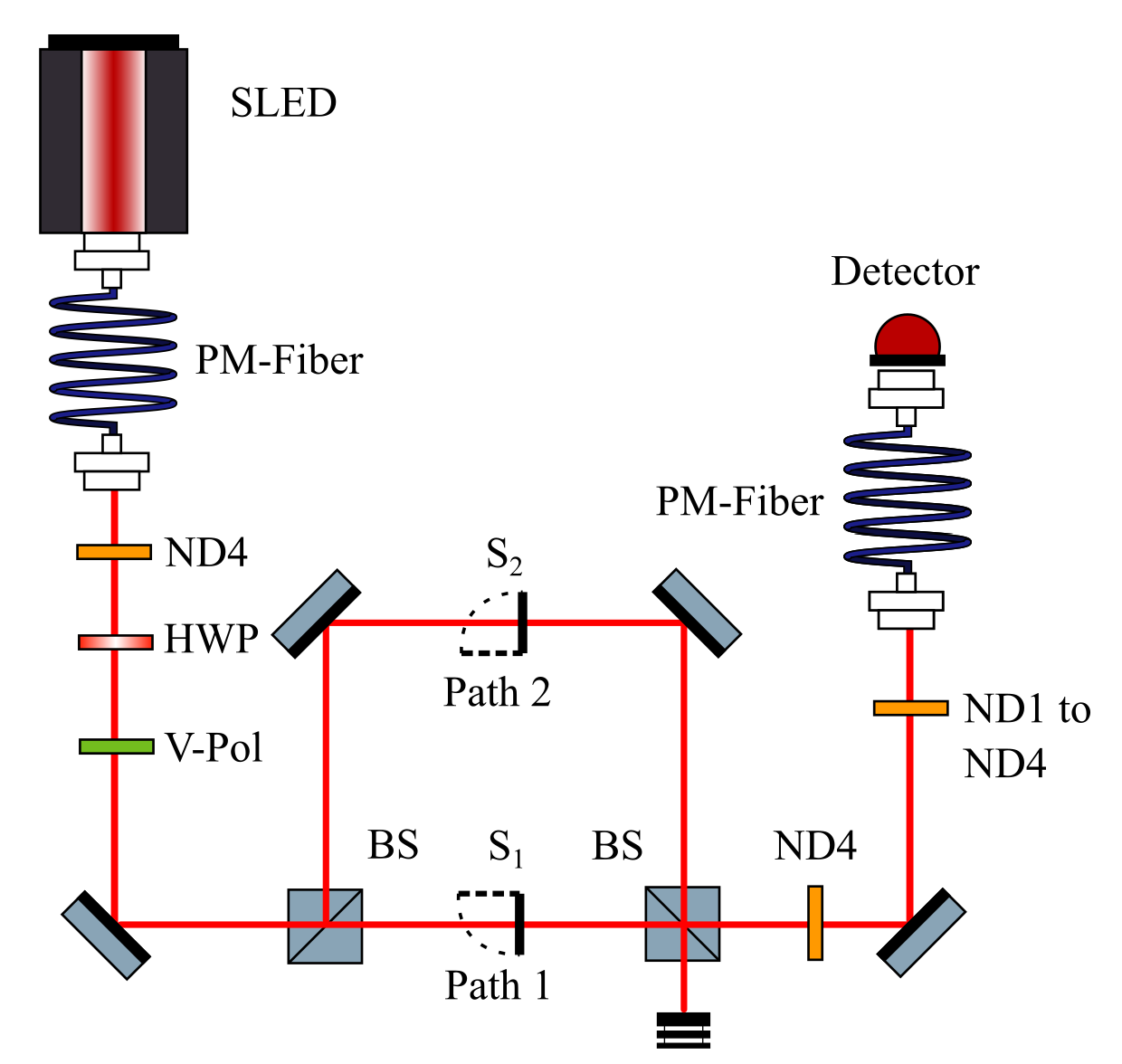}\label{fig_setup_c}}
    \caption{Linear vs. nonlinear detector response functions and schematic of the measurement setup. (a) A sublinear detector response (red) as arising via saturation effects. (b) A supralinear response (blue), as produced by multi-photon detection. (c) Layout of the experimental setup implementing the superposition method. To maintain constant polarisation, polarisation-maintaining (PM) fibers are used to couple light in and out of the setup. BS: beam splitter, HWP: half-wave plate. ND: Neutral density filter}
    \label{fig_setup}
\end{figure}

This experiment is based on the superposition method \cite{fast_nonlin_det_poly_Coslovi:80,kautenNonlin,Hlousek2023} and schematically illustrated in Fig.~\ref{fig_setup_c}. Light from a superluminescent diode (SLED; central wavelength \SI{1528}{\nm}, FWHM bandwidth \SI{90}{\nm}) is equally split into two paths $1$ and $2$, which can be individually blocked by shutters $S_{1(2)}$. One of the output modes of the recombining beam splitter is then fiber-coupled and detected by the device under test. In order to avoid any interference effects, the two beam paths have a length difference of $\Delta l=\SI{26}{\cm}$ much greater than the coherence length of the light source (\SI{26}{\micro m}). Then the photon fluxes of the two paths $\phi_{1(2)}$, which are incident on the detector, add up incoherently as $\phi_1+\phi_2 = \phi_{1+2}$, with $\phi_{1+2}$ denoting the combined photon flux for both shutters being open.\\

Let the detected photon flux be $\hat{\nu_i}=f(\phi_i)$, with the (unknown) detector transfer function $f$ and shutter setting $i=1,2,1+2$. We subtract the background $\nu_0$, i.e. the registered flux for both shutters being closed, from the raw measured signal to obtain the background-corrected detected flux 
$\nu_i = \hat{\nu_i} - \nu_0$. Following the superposition method, one can obtain the residuum $r$ \footnote{Note that the sign of the residuum $r$ in Eq.~\ref{eq_res_r} is here chosen opposite to Ref.~\onlinecite{kautenNonlin}.}
\begin{equation}\label{eq_res_r}
  r= \nu_1 + \nu_2 - \nu_{1+2},
\end{equation}
as a measure for the nonlinearity of the detector setup.
Additionally, Eq.~\eqref{eq_res_r} can be divided by $\nu_{1+2}$ to obtain a dimensionless expression $\Delta$ for the nonlinearity \cite{Hlousek2023}:
\begin{equation}\label{eq_res_delta}
  \Delta= \frac{r}{\nu_{1+2}} = \frac{\nu_1 + \nu_2}{\nu_{1+2}} - 1.
\end{equation}
A perfectly linear detector response would give $\Delta=0$ at all photon flux levels, whereas $\Delta>0$ and $\Delta<0$ would indicate a sub- and supralinear response, respectively.

To measure the nonlinearity for different values of the combined photon flux $\phi_{1+2}$, we use a half-wave plate (HWP) set at angle $\alpha_i$ and a linear polariser (see left side in Fig.~\ref{fig_setup_c})
to realise equidistant power steps $P_i=P_0 \cdot \cos^2(2\alpha_i)$.


The HWP is mounted on a motorised rotation stage to automatically set different power levels for the measurement. The two shutter mechanisms are realised via metal plates on motorised stages, which rotate the plates into and out of their respective beam path.
The polarisation of the fiber-coupled output signal is controlled via a 3-paddle fiber polarisation controller (not shown in the figure) maximising the detected signal of the polarisation-sensitive SNSPDs.


As detector we investigate a commercially available SNSPD with a NbTiN nanowire in a closed-cycle cryostat operating at \SI{2.9}{K} (Single Quantum EOS 720 CS) optimised for telecom wavelengths around \SI{1550}{\nm}, together with a time tagger device (Swabian instruments). Neutral density (ND) filters are used to attenuate the SLED signal to a power level compatible with the SNSPD, in particular keeping the detected count rate below \SI{600}{kHz} to prevent latching. The default bias current for this detector is $I_{\mathrm{b}} = \SI{22}{\uA}$, which is used in all measurements, unless otherwise stated. For each input flux level set by the HWP rotation angle, we measure the nonlinearity repeatedly over $60$ cycles of shutter settings ($i=0,1,2,1+2$) with an integration time of \SI{10}{\s} per setting. 

\section{Analysis of Results}\label{Analysis}
\begin{figure}[h]
    \includegraphics[width=0.85\textwidth]{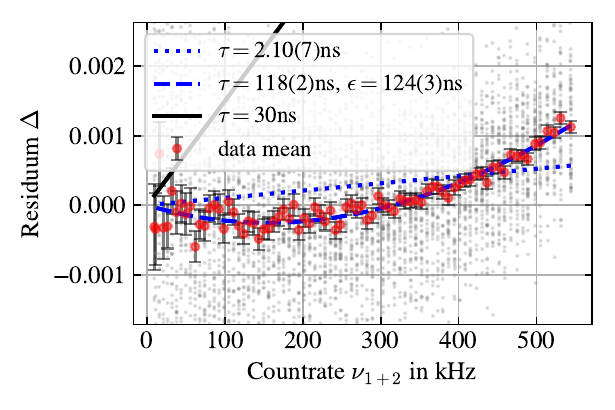}
    \caption{
    Nonlinearity measurement (Residuum $\Delta$) of the SNSPD for default detector settings. The gray dots indicate single measurements whereas the red dots show the average for each countrate. The uncertainty bars are given by the standard error of mean. The blue lines are two different analytic transfer function models [Eq.~\eqref{eq_deadtime_trans2}, dotted line, and Eq.~\eqref{eq_tpd_trans2}, dashed line] fitted and applied to the residuum. The black solid line shows the dead time model transfer function [Eq.~\eqref{eq_deadtime_trans2}] with a dead time of \SI{30}{\ns}.}
    \label{fig_Res_22}
\end{figure}

Within this section a detailed analysis of the experimental data is given. 
The results are displayed in Fig.~\ref{fig_Res_22} 
as a function of the measured background-corrected countrate with both paths open $\nu_{1+2}$. The grey dots show the obtained residual $\Delta$ for each of the $60$ cycles and demonstrate the spread of the individual measurements due to shot noise. The average for each countrate (red dots) reveals a systematic trend of these residuals, initially slightly towards negative values (where the sum of the single-path signals is smaller than the combined signal), reversing to positive values (vice versa) for countrates above \SI{300}{kHz}.
In the next subsection, we model the observed dynamics via the known nonlinear processes of detector dead time and two-photon detection.

\subsection{Analytic transfer function models}\label{results_analytic_models}

A commonly used model for describing the nonlinearity of a single-photon detector considers the dead time $\tau$ after successfull registration of a photon. Such dead times are naturally present in all avalanche detectors, but are also unavoidable in superconducting nanowires, as the recovery of the supercurrent after a detection event requires a finite time, which is specified by the manufacturer for our system as $\tau\approx\SI{30}{\ns}$. The dead time leads to an effective saturation of the detector and can, in its simplest form, be modelled by a step-like recovery, leading to a detector transfer function \cite{kautenNonlin}
\begin{equation}\label{eq_deadtime_trans2}
    \hat{\nu_i}=f(\phi)\approx\nu=\frac{\phi}{1+\tau\phi}.
\end{equation}
We make here the approximation that the background event rate $\nu_0$ is much smaller than the photon fluxes in the range of interest, such that their contribution to the detector saturation can be neglected. This assumption is well satisfied in our case, where $\nu_0\lesssim\SI{350}{\Hz}$ (arising mostly from detector dark counts) and $\phi$ is in the order of tens to hundreds of \si{\kHz}. We have also tested more elaborate transfer functions taking into account the background contribution, similar to Eq.~(2) in Ref.~\onlinecite{Hlousek2023}, but found no significant influence on the fits to the experimental data of Fig.~\ref{fig_Res_22}.
Assuming equal photon fluxes from both paths $\phi_1=\phi_2$ and $\phi_1+\phi_2=\phi_{1+2}$ (i.e. the fluxes do superimpose in reality) leads to the following explicit function for the normalised residuum (see appendix \ref{sec:AppTransFunc} for a derivation) 
\begin{equation}\label{eq_fit_deadtime}
\Delta = \frac{2}{2-\tau\nu_{1+2}}-1.
\end{equation}
We use this equation to fit the dead time model to the measured data as shown in Fig.~\ref{fig_Res_22} (blue dotted line).
This theoretical prediction does not match our data and, thus, the fitted parameter $\tau= \SI{2.10(7)}{\ns}$ cannot be representative for the real dead time in our system. Also a dead time model with the expected parameter $\tau=\SI{30}{\ns}$ (black line) does not match our data. In fact, it would predict a much steeper rise of the residuum. This implies that in our system the observed nonlinearity is much smaller than expected from a dead time in the order of tens of \si{ns}. This suggests that the count rate saturation due to detector dead time is counteracted by another nonlinear effect with opposite sign. 
\\

One explanation for a second nonlinear phenomenon is the occurrence of \ac{tpd} events \cite{P_click}. Empirically, this effect can be modelled as 
\begin{equation}\label{eq_tpd_trans}
    f(\phi)\approx\nu = \phi + \epsilon \phi^2,
\end{equation}
where again the detector saturation due to dark counts is neglected and $\epsilon$ describes an effective time interval within which two incident photons trigger a detection. This effect on its own would lead to purely negative values for $\Delta$. Combining the two transfer functions of the dead time model Eq.~\eqref{eq_deadtime_trans2} and the \ac{tpd} model Eq.~\eqref{eq_tpd_trans} results in a new transfer function
\begin{equation}\label{eq_tpd_trans2}
    f(\phi)\approx\nu = \left( \frac{\phi}{1+\tau  \phi} \right)+\epsilon \left(\frac{\phi}{1+\tau \phi}\right)^2,
\end{equation}
The resulting fit function (see appendix \ref{sec:AppTransFunc} for an analytic expression) for the residuum together with its fitting parameters $\tau$ and $\epsilon$ is shown in Fig.~\ref{fig_Res_22} as a blue dashed line. 
This combined dead time-\ac{tpd} model fits the measured data well with a normalised $\chi^2 = \num{0.81}$. This may be an indication for the presence of \ac{tpd}-related effects. Despite the well matching model, the fitted dead time $\tau=\SI{118}{\ns}$ is significantly higher than expected. This leads to the conclusion that neither of these basic analytic models delivers a realistic representation of the detector physics. In order to develop a more substantiated understanding, we first corroborate the influence of two-photon detections on the nonlinearity via a variation of the bias current supplied to the nanowire (see Sec.~\ref{results_bias_current_dep}), then directly measure the recovery of the detector efficiency (Sec.~\ref{results_direct_meas_recovery}), and finally simulate the detection statistics taking all these observations into account (Sec.~\ref{results_sim_SDE}).

\subsection{Bias current dependence}\label{results_bias_current_dep}

\begin{figure}[h]
    \includegraphics[width=0.95\textwidth]{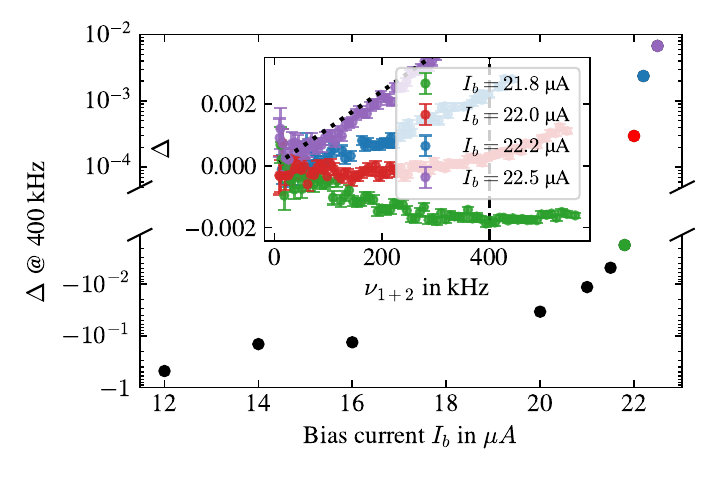}
    \caption{Extracted nonlinearity at a given countrate from measurements at different bias current values. The red point indicates the weakest NL for this countrate with $I_{\mathrm{b}} = \SI{22}{\uA}$. 
    The insert plot shows residuum curves with different bias currents. The measurement with $I_{\mathrm{b}} = \SI{22.5}{\uA}$ (purple dots) is fitted with the deadtime model of equation Eq.~\eqref{eq_deadtime_trans2} (dotted line).}
    \label{fig_Res_diff_ib}
\end{figure}

The balance between single-photon and two-photon detection events in an SNSPD depends strongly on the applied bias current \cite{P_click}. In particular, a decrease of bias current is known to increase the relative contribution of \ac{tpd}-events, which in turn should lead to a more pronounced negative trend in the residuum $\Delta$.
We perform several measurements with bias current settings in the range from \SI{12}{\uA} to \SI{22.5}{\uA}. 
To compare these measurements, we plot one residuum value $\Delta(\SI{400}{\kHz})$ for each bias current at the same countrate (see Fig.~\ref{fig_Res_diff_ib}) \footnote{A $5^{(\text{th})}$-order polynomial interpolation of the data was used to extract $\Delta$ at exactly the same countrate for all bias currents.}. 

An insert plot shows several nonlinearity measurements, around the default bias current ($I_{\mathrm{b}}=\SI{22}{\uA}$, i.e red dots, corresponding to the average data in Fig.~\ref{fig_Res_22}).
One can observe a transition from a purely positive residual ($I_{\mathrm{b}}=\SI{22.2}{\uA}$, blue dots) to a negative one ($I_{\mathrm{b}}=\SI{21.8}{\uA}$, green dots). For even lower bias currents, the nonlinearity becomes even more negative, which matches the expectation of an increasing contribution from \ac{tpd} events. In the opposite direction, for increasing bias currents, the \ac{tpd} contribution should diminish and the dead time effects dominate. A measurement with $I_{\mathrm{b}}=\SI{22.5}{\uA}$ shown in Fig.~\ref{fig_Res_diff_ib}, fitted with the dead time model with $\tau = \SI{23.3(1)}{\ns}$ (dotted line), which approaches the manufacturer specification of $\SI{30}{\ns}$, enforces this observation. The nonlinearity with smallest magnitude at this particular countrate indeed occurs at the default bias current ($I_{\mathrm{b}} = \SI{22}{\uA}$). \\



\subsection{Direct measurement of recovery time}\label{results_direct_meas_recovery}

\begin{figure}[htb!]

\sidesubfloat[]{%
 \includegraphics[width=0.45\textwidth]{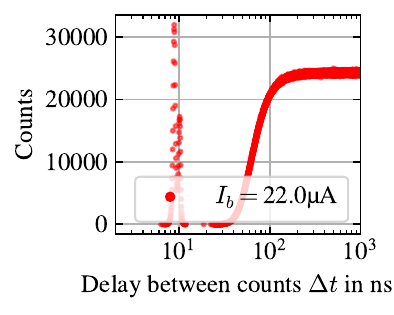}\label{fig_histogram_raw}}
\sidesubfloat[]{%
\includegraphics[width=0.42\textwidth]{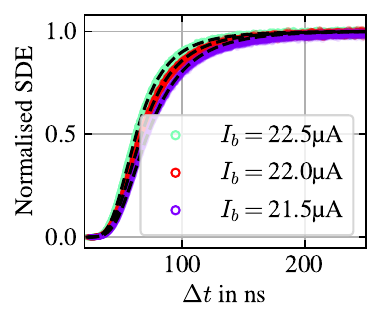}\label{fig_histogramSDE}}
        \caption{Start-stop histogram measurements. Count rates are within the range \SIrange{300}{400}{kHz} for all $I_{\mathrm{b}}$. (a) Raw data for a single bias current value. The first initial peak is the result of noise in the voltage spike of the detector around the trigger voltage. Thus, counts below \SI{20}{\ns} delay are ignored. (b) Normalised histogram (scaled to 1 for counts with $\Delta t=\SI{800}{\ns}$ and counts with $\Delta t<\SI{20}{\ns}$ are removed). This can be interpreted as a normalised system detection efficiency. All three data sets are fitted with \eqref{eq_SDE(dt)} and the resulting fit parameters are given in Table~\ref{tab_histogram_fit}. The fitted curves are shown as dashed lines.}
    \label{fig_histogram}
\end{figure}

 One key limitation of the dead time model [Eqs.~\eqref{eq_deadtime_trans2} and ~\eqref{eq_tpd_trans2}] is the assumed step-like detector recovery, while in reality the detection efficiency reaches its nominal values only gradually after a photon detection \cite{DirectMeasOfRecoveryTime}. This results in incorrect dead time estimations. To overcome this limitation we directly measure the detector recovery via the autocorrelation of detection times. Specifically, we measure the distribution of time-differences $\Delta t$ between all (not only subsequent) detection events via a multiple start - multiple stop - histogram measurement (see Fig.~\ref{fig_histogram_raw}). The first peak at about \SI{9}{\ns} in this plot arises from noise in the voltage pulse of a detection signal\footnote{a noise spike in the trailing edge of the pulse can trigger the counting electronics a second time} and, thus, can be disregarded. The distribution of the remaining clicks is directly proportional to the temporal recovery of the \ac{sde}. Evidently, there are no time differences smaller than about \SI{30}{\ns} visible in the data, in accordance to the dead time specified by the manufacturer. However, the efficiency does only gradually recover, with a \SI{90}{\percent}-recovery in the order of \SI{100}{\ns} (see normalised curve in Fig.~\ref{fig_histogramSDE}). We followed the procedure described in Ref.~\onlinecite{DirectMeasOfRecoveryTime} to fit the \ac{sde} $\eta$
 as a function of $\Delta t$. This requires two steps: 

First we determine the dependence of the (steady-state) \ac{sde} on the bias current via measuring the countrate as a function of $I_\mathrm{b}$, see appendix \ref{sec:AppSDE}. This dependence can be modelled with an error function
 \begin{equation}\label{eq_SDE(dt)}
\eta(I_{\mathrm{b}}) = 0.5 \eta_{\mathrm{max}}\left(1+\erf{ \left(\frac{I_{\mathrm{b}}-I_0}{\Delta I}\right)}\right),
\end{equation} 
with parameters $\eta_{\mathrm{max}}=0.680(1)$, $I_0=\SI{19.962(4)}{\uA}$ and $\Delta I=\SI{2.343(7)}{\uA}$. In a second step, the recovery dynamics of the \ac{sde} $\eta(\Delta t)$ is modelled by substituting the steady-state $I_{\mathrm{b}}$ in Eq.~\eqref{eq_SDE(dt)} by a time-dependent bias current $I(\Delta t)$, with the following temporal profile:
\begin{equation}\label{eq_bias_dynamics}
I(\Delta t) = \left( I_{\mathrm{b}} - I_{\text{drop}}\right) \left(1-e^{-\Delta t/\tau}\right) + I_{\text{drop}},
\end{equation}
with fit parameters $I_{\text{drop}}$ and $\tau$.
This profile follows from the analog circuit model of the SNSPD \cite{DirectMeasOfRecoveryTime}: 
After a detection, the current drops very rapidly from the steady-state bias current $I_{\text{b}}$ to a lower current $I_{\text{drop}}$ (enabling the nanowire to cool down, also very rapidly). As soon as the superconductivity gets restored, the current rises again with a characteristic time $\tau$ (set by the load resistance of the readout electronics and the kinetic inductance of the nanowire) and, thus, also the detection efficiency rises according to Eq.~\eqref{eq_SDE(dt)}. Fig.~\ref{fig_histogramSDE} shows the fitted \ac{sde}, normalised to the efficiency at infinite delay, as dashed lines for different $I_{\mathrm{b}}$, which agrees well with the data. The resulting fit parameters are given in Table~\ref{tab_histogram_fit}.

\begin{table}
\caption{\label{tab_histogram_fit}Fit parameters of Fig.~\ref{fig_histogramSDE} and measured $90\%$-recovery time $t_{0.9}$.}
\begin{ruledtabular}
\begin{tabular}{ccccc}
 $I_{\mathrm{b}}$  &   $\tau$  & $I_{\text{drop}}$ & $\chi^2/\text{d.o.f.}$ & $t_{0.9}$ \\ 
  \midrule
 $\SI{22.5}{\uA}$ &   $\SI{33(4)}{\ns}$  & $\SI{5.23(1)}{\uA}$ & \num{1.49} & \SI{99}{\ns}\\  
 $\SI{22.0}{\uA}$  &  $\SI{34(4)}{\ns}$  & $\SI{5.57(1)}{\uA}$ & \num{1.77} & \SI{109}{\ns}\\  
 $\SI{21.5}{\uA}$  &  $\SI{34(4)}{\ns}$  & $\SI{6.06(2)}{\uA}$ & \num{2.12} & \SI{121}{\ns}\\
\end{tabular}
\end{ruledtabular}
\end{table}

With this method one can define and determine the dead time of the detector as the recovery time constant $\tau$, which matches well with the specified value of $\SI{30}{\ns}$ for all investigated bias currents. Fig.~\ref{fig_histogramSDE} further shows that the time required for a full recovery of the detector increases for lower bias current values, see also the \SI{90}{\percent}-recovery times listed in the rightmost column in Table~\ref{tab_histogram_fit}. This is due to the steepest part of the SDE vs. bias current curve being located around $I_0\approx\SI{20}{\uA}$ (see appendix \ref{sec:AppSDE}), which is reached later for lower steady-state bias currents.
In the context of nonlinearity, such changes in the total recovery time also have an impact on the residuum. Lowering $I_{\mathrm{b}}$ increases the effective recovery time, therefore strengthening the sublinear contribution and leading to an increase in $\Delta$. In practice, however, Fig.~\ref{fig_Res_diff_ib} shows the opposite trend. This indicates that other effects influencing the nonlinearity, like \ac{tpd} as discussed in the previous section, must exhibit a stronger bias current-dependence than the recovery time.\\

The fitted model of the detector-recovery can now be used to predict the detection probability of subsequent photons in the context of a Monte-Carlo simulation, which will the starting point of the next section.

\subsection{Statistical simulation of detector response}\label{results_sim_SDE}

We now aim to gain some insight into the physics underlying the \ac{tpd} and its contribution to the detector nonlinearity via a statistical modelling of the photon absorption processes and recovery dynamics. To this end, we perform a Monte-Carlo simulation with randomly distributed photon arrivals, taking into account the measured detector efficiency recovery $\eta(\Delta t)$ from the previous section \cite{DirectMeasOfRecoveryTime}. In a second step, we incorporate to the model the formation and decay of resistive hotspots in the nanowire \cite{P_click}. In particular, we consider 
a sequence of $N+1$ photons with a given incidence rate $\phi$ and Poissonian count statistics with exponentially distributed arrival time differences $\Delta t_i$ , $i \in \{1,...,N\}$. Note that we always start with a photon detection at $t=0$, which starts the clock for the detector recovery. The first random photon ($i=1$) then arrives at the detector at time $\Delta t_1$. For each photon $i$ the detection probability $p_i$ can be calculated as
\begin{equation}
    p_i = A\eta(\Delta t_i +\delta t),
    \label{eq:MCLinearProb}
\end{equation}
where $A=0.604$ is the measured steady-state SDE of the detector and $\delta t$ is the time between the last successfully detected photon and the arrival of the previous photon $i-1$, i.e. it accounts for the continued recovery if a photon is not detected. By definition $\delta t=0$, if photon $i-1$ was detected.
If then photon $i$ does not trigger a click, the following photon has the detection probability $p_{i+1}=A\eta(\Delta t_i + \Delta t_{i+1})$. In contrast, the recovery starts anew, if photon $i$ is detected ($p_{i+1}=A\eta(\Delta t_{i+1})$). This Monte-Carlo simulation can be repeated for different photon rates $\phi_{1+2}^{\text{sim}}$ and respectively $\phi_{1(2)}^{\text{sim}}=\phi_{1+2}^{\text{sim}}/2$ to simulate the single- and two-path measurements of the experiment. The outcome of these simulations is then used to calculate the simulated residuum
\begin{equation}\label{eq_delta_sim}
\Delta^{\text{sim}} = \frac{2 \nu_{1,2}^{\text{sim}}}{\nu_{1+2}^{\text{sim}}}-1.
\end{equation}
Running this Monte-Carlo simulation as described always yields a purely positive nonlinearity $\Delta$, similar to the standard dead time model.
\\

To incorporate the effect of \ac{tpd}, we consider the hotspot relaxation dynamics in an SNSPD. A photoabsorption in the superconducting nanowire leads to a local breakdown of Cooper pairs, inducing a resistive hotspot. Consequently, the current concentrates in the remaining superconducting regions, surpassing the critical current density and resulting in a growing hotspot. This growth is counteracted by cooling. Depending on the balance between the two processes, either a resistive barrier forms over the entire cross-section of the nanowire, resulting in a sudden, detectable voltage rise, or the hotspot diminishes over time and nothing is detected. But even in the latter case the presence of the hotspot will render the nanowire more likely to transition to the normal conducting state when a subsequent photon is absorbed. For sufficiently low bias currents, the system can even be forced into a \ac{tpd} regime, where single absorptions are almost never detected, while two subsequent absorptions can trigger a detection \cite{P_click}. 
To model this process, we modify the probability function of Eq.~\eqref{eq:MCLinearProb} as
\begin{equation}
    p_i= A\eta(\Delta t_i+\delta t) + B_i,
\end{equation}
where 
$B_i$ implements a boost term for the detection of photon $i$ that accounts for temporally and spatially overlapping hotspots. If a single incident photon does not trigger a detection in the simulation, the detection probability of the following photon (within the mean hotspot lifetime) gets increased by the boost term.

\begin{figure}[h]
\sidesubfloat[]{%
    \includegraphics[width=0.7\textwidth]{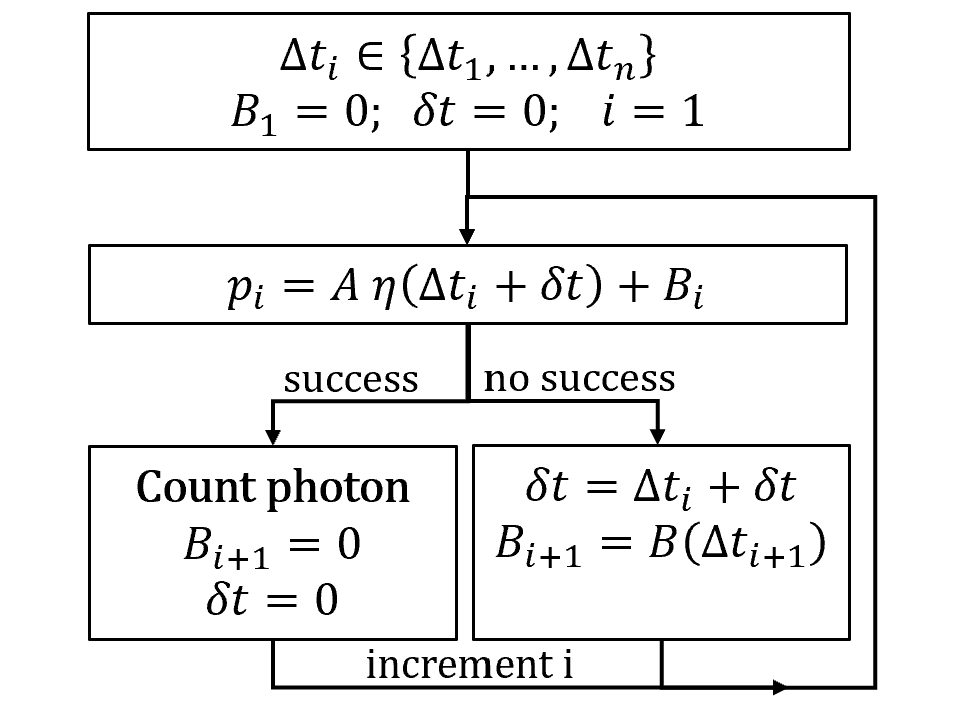} \label{fig_sim_concept_a}}
    \hfill
\sidesubfloat[]{%
    \includegraphics[width=0.7\textwidth]{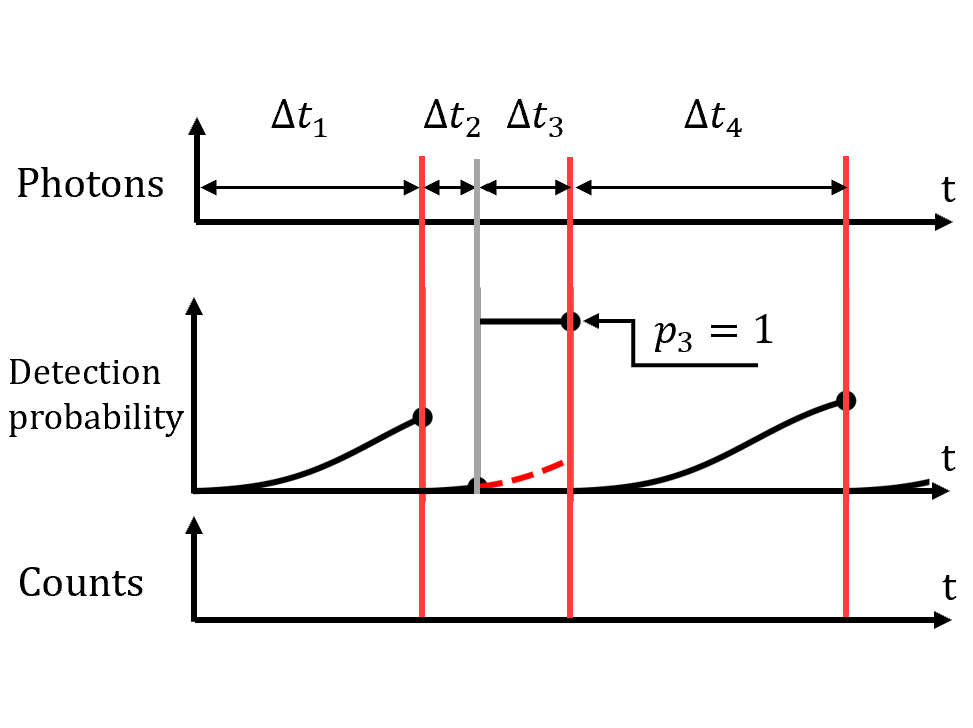} \label{fig_sim_concept_b}}
    
    \caption{(a) Flowchart of the simulation concept. (b) Exemplary timeline of random photon arrivals. Photons are either detected (red vertical lines), restarting the detector recovery from zero or not detected (grey line), causing a boost of the detection probability from the standard recovery (red dashed curve), here to unity, until the next photon is detected or a maximal time $T_{\mathrm{boost}}$ is reached.}
    \label{fig_sim_concept}
\end{figure}

Fig.~\ref{fig_sim_concept_a} shows a block diagram of the simulation. The boost term is implemented by
\begin{equation}
  B_{i+1}=\begin{cases}
    B_0\cdot b(\Delta t_{i+1}, T_{\text{boost}}) & \text{if $i$ not successful},\\
    0 & \text{otherwise},
  \end{cases}
\end{equation}
with amplitude $B_0$, time evolution $b(\Delta t_{i+1},T_{\text{boost}})$ and characteristic time scale of the boost process $T_{\text{boost}}$. For the sake of simplicity we assume that each simulated photon hits the nanowire, creates a hotspot and results in either a click or a boost for the next photon within the characteristic time $T_{\text{boost}}$. Therefore, this procedure can serve as an upper-bound estimate of the influence of \ac{tpd} effects.\\

\begin{figure}[htb!]
    \includegraphics[width=0.9\textwidth]{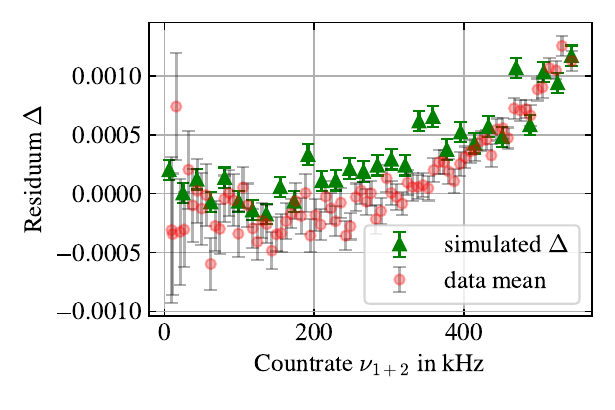}
    \caption{Simulated detector residuum with boost implementation accounting for the hotspot relaxation dynamics in the nanowire. Simulation parameters are $N=\num{2e8}$, $T_{\text{boost}}=\SI{172(1)}{\ns}$ and $A = \num{0.604}$.}
    \label{fig_res_sim_main}
\end{figure}

The simplest
implementation for the boost is a step-function, that increases the detection probability for subsequent photons to one. This implies that after an unsuccessful detection the next photon arriving within $T_{\text{boost}}$ will be detected with certainty. We can realise this scenario with $B_0=1-A\eta(\Delta t_i+\delta t)$ and $b(\Delta t_{i+1})=u(T_{\text{boost}}-\Delta t_{i+1})$, with step-function $u(x)=1$ for $x>0$ and $u(x)=0$, otherwise.
We repeated the simulation for various values of the characteristic time $T_{\text{boost}}$. The simulation result, which optimally fits to the measured data is shown in Fig.~\ref{fig_res_sim_main}, where $T_{\text{boost}} = \SI{172(1)}{\ns}$. Note that the uncertainty of $T_{\text{boost}}$ is the variation stepsize.

The simulated residuum of Fig.~\ref{fig_res_sim_main} does not match the data exactly but lies within the same range. The physical interpretation of this simulation result is, however, rather surprising, as the required time constant $T_{\text{boost}}$ is much larger than the known hotspot relaxation times in the \ac{tpd} regime, i.e., at low bias currents, which are in the order of \SI{100}{\ps} and below \cite{Ferrari:17_hs_relax_time}. 
A further issue of this simulation is the 
assumption that all photons get absorbed and the boost effect is maximal, which is certainly an overestimation. Making more realistic assumptions would result in even longer $T_{\text{boost}}$. On the other hand, no experimental data for the hotspot relaxation in the single-photon detection regime, in which our system operates at the default $I_{\text{b}}$, are available, and the \ac{tpd}-data shows a clear trend towards longer relaxation times for increasing bias currents \cite{P_click}. \\

\section{Discussion}
Our experimental results clearly show that the relative nonlinearity of the SNSPD under test, as measured by the residuum $\Delta$ is below $\left|\Delta\right|=\num{1e-3}$ for count rates up to \SI{500}{kHz}, when the detector is operated at its default bias current of $I_{\mathrm{b}}=\SI{22}{\uA}$. From a dead-time induced detector saturation alone, however, much larger (sublinear) nonlinearities would be expected. This indicates the presence of another, (supralinear) counteracting effect, which can be empirically modelled via TPD-events. The agreement between the data and the empirical model is very good (cf. dashed curve in Fig.~\ref{fig_Res_22}), such that this model can in principle be used to compensate the nonlinearity bias in comparative measurements \cite{Kauten_2017_5Path}. The balance between the nonlinear contributions of dead time and TPD can be adjusted by varying the detector bias current, with dead time dominating at larger currents and TPD at lower ones. Remarkably, the optimal balance with minimal nonlinearity at typical fluxes of a few \SI{100}{\kHz}, occurs here indeed at the default setting, which is recommended by the manufacturer to achieve a good compromise between detector efficiency and dark counts.

Our statistical simulations, which took into account the measured detector recovery and modelled the TPD via the generation and relaxation of resistive hot spots in the nanowire, could reproduce the observed nonlinearity data. Only the resulting time scales of hot spot relaxation seem unexpectedly large and call for a deeper investigation of their physical origin. A definite determination of the time scale would also be important to correctly set the repetition rate of a single-photon source, if one wishes to avoid nonlinearities altogether \cite{Vogl2021}. 

As one possible cause for the apparent large $T_{\mathrm{boost}}$, we considered the effect of count-rate dependent detection efficiency, known as AC-biasing \cite{AC_biasing}. This effect arises in SNSPDs due to charges from a detection pulse accumulating at a capacitor in the AC-coupled readout electronics. As a consequence, the effective bias current through the nanowire, and with it the detection efficiency, increase for increasing count rates \cite{Ferrari:19_detect_response_AC_biasing}, giving rise to a supralinear dynamics and decreasing the residuum $\Delta$. We estimated the impact of this effect by repeating the measurement of the detector recovery (cf. Sec.~\ref{results_direct_meas_recovery}) at two different flux levels and incorporating the measured efficiency curves into the Monte-Carlo simulation at these particular levels. The observed reduction of $\Delta$ yields a lower optimal $T_{\text{boost}}=\SI{161}{\ns}$, that is a reduction of the hot-spot relaxation time by approximately \SI{6}{\percent} (see appendix \ref{sec:ACBias}). This is consistent with the AC-biasing contribution to $\Delta$ being one order of magnitude smaller than the individual contributions of the two main effects of dead time and TPD. We further considered photon bunching on short time scales arising from the thermal photon number statistics of the SLED \cite{SLED_g2}. However, this had a negligible impact on the simulation due to the very short coherence time of the light source ($<\SI{100}{\fs}$).

From a broader perspective, our results highlight the necessity to precisely and systematically characterise photon detector nonlinearities, as they can influence the results of comparative measurements on the $10^{-3}$-level of relative accuracy, even when the dynamic range of the measurement is only at the modest factor of two. For SNSPDs, the results suggest that the amount and sign of this nonlinearity can crucially depend on the applied bias current, so a careful calibration might be needed for each individual detector and its setting. It would be very interesting to develop a deeper understanding of the multi-photon detection contribution to the nonlinearity and its involved time scale to allow the formulation of a quantitative predictive model.

\begin{acknowledgments}
The authors thank Simone Ferrari, Josef Hlou\v{s}ek and Philipp Zolotov for fruitful discussions and Stefanie Morhenn for help in setting up the experiment. This research was funded by the Austrian Science Fund (FWF), grants no. 10.55776/FG 5, 10.55776/F71 and 10.55776/P30459.
\end{acknowledgments}

\section*{Data Availability Statement}
The data that support the findings of this study are openly available in Zenodo at \url{https://doi.org/10.5281/zenodo.12773197}.

\appendix

\section{Analytic transfer functions}
\label{sec:AppTransFunc}
For the basic dead time model, the detector transfer function of Eq.~(\ref{eq_deadtime_trans2}) can be inverted to
\begin{equation}\label{eq_deadtime_trans_inv}
    \phi=f^{-1}(\nu)=\frac{\nu}{1-\nu\tau}.
\end{equation}
Under the assumptions of equal fluxes along both paths and absence of interference, i.e., $\phi_1=\phi_2=\phi_{1+2}/2$, which implies also $\nu_1=\nu_2$, one obtains for the normalised residuum
\begin{equation}\label{eq_res_deadtime_inv}
    \Delta=\frac{\nu_1+\nu_2}{\nu_{1+2}}-1=\frac{2\nu_1}{\nu_{1+2}}-1=\frac{2f\left[\frac{1}{2}f^{-1}(\nu_{1+2})\right]}{\nu_{1+2}}-1.
\end{equation}
Substituting Eq.~(\ref{eq_deadtime_trans_inv}) into~(\ref{eq_res_deadtime_inv}) then yields the fit function of Eq.~(\ref{eq_fit_deadtime}) from the main text.\\

In an analog fashion, the transfer function Eq.~\ref{eq_tpd_trans2} combining dead time saturation and \ac{tpd} can be inverted to
\begin{equation}
    \phi=f^{-1}(\nu)=\frac{2\nu}{1+\sqrt{1+4\epsilon\nu}-2\nu\tau}.
\end{equation}
Substituting this expression into Eq.~(\ref{eq_res_deadtime_inv}) yields the following analytic expression for the residuum:
\begin{eqnarray*}
   \Delta(\nu)= -1+\frac{2(-1+\sqrt{1+4\epsilon\nu}+2\tau\nu)}{\nu(4\epsilon+\tau(3+\sqrt{1+4\epsilon\nu}-2\tau\nu))^2} \times \nonumber\\
   \left[\tau(3+\sqrt{1+4\epsilon\nu}-2\tau\nu)+\epsilon(3+\sqrt{1+4\epsilon\nu}+2\tau\nu)\right],
\end{eqnarray*}
which is used as a fit function in Fig.~\ref{fig_Res_22} with $\nu=\nu_{1+2}$ and fit parameters $\tau$ and $\epsilon$.

\section{System detection efficiency}
\label{sec:AppSDE}
\begin{figure}[htb!]
    \includegraphics[width=0.8\textwidth]{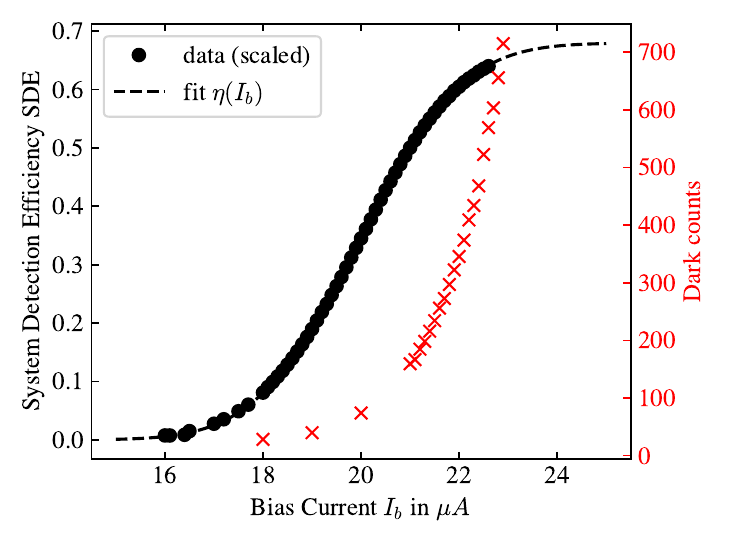}
    \caption{System Detection Efficiency (\ac{sde}) as a function of bias current $I_b$. Additionally, dark counts in units of $\si{s^{-1}}$ are plotted in red. Fit parameters of Eq. \eqref{eq_SDE(dt)} in the main text are $I_0=\SI{19.962(4)}{\uA}$, $\Delta I=\SI{2.343(7)}{\uA}$ and $\eta_{max}=0.680(1)$.}
    \label{fig_SDE_vs_Ib}
\end{figure}

We measured the count rate for different values of the bias current $I_{\mathrm{b}}$ at a detected flux of about $\SI{300}{\kHz}$ at $I_{\mathrm{b}}=\SI{22}{\uA}$, see Fig.~\ref{fig_SDE_vs_Ib}. The data was scaled to the manufacturer-specified efficiency of $0.64$ at the largest possible bias current $I_{\mathrm{b}}=\SI{22.6}{\uA}$ before the count rate dropped due to latching. Note that the specific value of this scaling constant is irrelevant, as the \ac{sde} curves in Fig.~\ref{fig_histogramSDE} in the main text are normalised to the count rate at infinite delay.

\section{AC-Biasing}
\label{sec:ACBias}

\begin{figure}[htb!]
    \sidesubfloat[]{
    \includegraphics[width=0.42\textwidth]{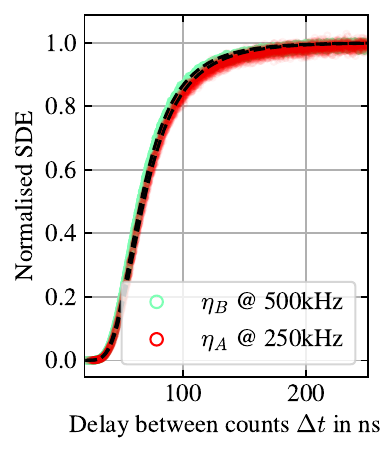}\label{fig_sde_ac_biasing}}
    \sidesubfloat[]{
    \includegraphics[width=0.46\textwidth]{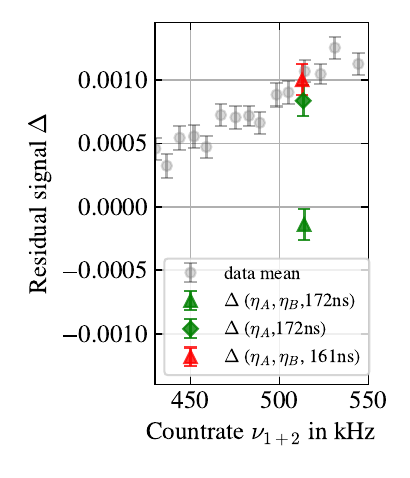}\label{fig_sim_ac_biasing}}
    \caption{Impact of AC-biasing. (a) Normalised start-stop histogram at $I_\text{b}=\SI{22}{\uA}$ for two different flux levels $\nu_\text{A}$ and $\nu_\text{B}$. (b) Simulated detector residuum at $\nu_{1+2}=\nu_\text{B}\approx\SI{500}{\kHz}$ without the AC-biasing effect (green diamond, equivalent to the simulation in Fig.~\ref{fig_res_sim_main}) and with AC-biasing (green triangle), both at $T_\text{boost}=\SI{172}{\ns}$. Decreasing the boost time to $T_{\mathrm{boost}}=\SI{161}{\ns}$ for the model with AC-biasing then yields the best fit to the experimental data (red triangle).}
    \label{fig_ac_biasing}
\end{figure}

To estimate the magnitude of AC-biasing in our detector, we investigate the detector recovery at the default $I_{\mathrm{b}}=\SI{22}{\uA}$ via start-stop histogram measurements (see Sec.~\ref{results_direct_meas_recovery}) at two different flux levels: $\nu_\text{A}\approx\SI{250}{\kHz}$ and $\nu_\text{B}\approx\SI{500}{\kHz}$. The measured recovery curves $\eta_\text{A}$ and $\eta_\text{B}$ are plotted in Fig.~\ref{fig_sde_ac_biasing} and the resulting fit parameters are given in Table~\ref{tab_histogram_acbias}. As expected for AC-biasing, a faster recovery is observed at higher flux levels, thus raising the average detection efficiency. We then repeat the Monte-Carlo simulation of Sec.~\ref{results_sim_SDE} for the fixed count rates $\nu_{1+2}=\nu_\text{B}$ and $\nu_1\approx\nu_2\approx\nu_{1+2}/2=\nu_\text{A}$, using the recovery curves of Fig.~\ref{fig_sde_ac_biasing} for each of them. This leads to a lower residuum $\Delta$, due to the AC-biasing having a larger influence on $\nu_{1+2}$ than on the lower rates $\nu_1=\nu_2$, counteracting the deadtime-induced nonlinearity. Repeating the simulation with $T_{\mathrm{boost}}=\SI{172}{\ns}$, as in Sec.~\ref{results_sim_SDE}, yields a reduction in $\Delta$ by about $0.001$ in comparison to a simulation without the AC-biasing effect, i.e., with a flux-independent recovery curve, see the green data points in Fig.~\ref{fig_sim_ac_biasing}. The best match of the simulation with AC-biasing to the experimental data (at $\nu_{\mathrm{1+2}}=\SI{500}{\kHz}$) is instead obtained for a slightly lower hotspot relaxation time $T_{\mathrm{boost}}=\SI{161}{\ns}$ (red data point).

\begin{table}
\caption{\label{tab_histogram_acbias}\ac{sde} fit parameters at different flux levels.}
\begin{ruledtabular}
\begin{tabular}{ccccc}
 $\nu$  &   $\tau$  & $I_{\text{drop}}$ & $\chi^2/\text{d.o.f.}$ & $t_{0.9}$ \\ 
  \midrule
 $\SI{250}{\kHz}$ &   $\SI{35(4)}{\ns}$  & $\SI{6.09(2)}{\uA}$ & \num{1.43} & \SI{112.5}{\ns} \\  
 $\SI{500}{\kHz}$  &  $\SI{33(4)}{\ns}$  & $\SI{5.256(9)}{\uA}$ & \num{1.97} & \SI{108.0}{\ns} \\  
\end{tabular}
\end{ruledtabular}
\end{table}

\bibliography{bibliography}

\begin{thebibliography}{38}%
\makeatletter
\providecommand \@ifxundefined [1]{%
 \@ifx{#1\undefined}
}%
\providecommand \@ifnum [1]{%
 \ifnum #1\expandafter \@firstoftwo
 \else \expandafter \@secondoftwo
 \fi
}%
\providecommand \@ifx [1]{%
 \ifx #1\expandafter \@firstoftwo
 \else \expandafter \@secondoftwo
 \fi
}%
\providecommand \natexlab [1]{#1}%
\providecommand \enquote  [1]{``#1''}%
\providecommand \bibnamefont  [1]{#1}%
\providecommand \bibfnamefont [1]{#1}%
\providecommand \citenamefont [1]{#1}%
\providecommand \href@noop [0]{\@secondoftwo}%
\providecommand \href [0]{\begingroup \@sanitize@url \@href}%
\providecommand \@href[1]{\@@startlink{#1}\@@href}%
\providecommand \@@href[1]{\endgroup#1\@@endlink}%
\providecommand \@sanitize@url [0]{\catcode `\\12\catcode `\$12\catcode `\&12\catcode `\#12\catcode `\^12\catcode `\_12\catcode `\%12\relax}%
\providecommand \@@startlink[1]{}%
\providecommand \@@endlink[0]{}%
\providecommand \url  [0]{\begingroup\@sanitize@url \@url }%
\providecommand \@url [1]{\endgroup\@href {#1}{\urlprefix }}%
\providecommand \urlprefix  [0]{URL }%
\providecommand \Eprint [0]{\href }%
\providecommand \doibase [0]{http://dx.doi.org/}%
\providecommand \selectlanguage [0]{\@gobble}%
\providecommand \bibinfo  [0]{\@secondoftwo}%
\providecommand \bibfield  [0]{\@secondoftwo}%
\providecommand \translation [1]{[#1]}%
\providecommand \BibitemOpen [0]{}%
\providecommand \bibitemStop [0]{}%
\providecommand \bibitemNoStop [0]{.\EOS\space}%
\providecommand \EOS [0]{\spacefactor3000\relax}%
\providecommand \BibitemShut  [1]{\csname bibitem#1\endcsname}%
\let\auto@bib@innerbib\@empty
\bibitem [{\citenamefont {Christensen}\ \emph {et~al.}(2013)\citenamefont {Christensen}, \citenamefont {McCusker}, \citenamefont {Altepeter}, \citenamefont {Calkins}, \citenamefont {Gerrits}, \citenamefont {Lita}, \citenamefont {Miller}, \citenamefont {Shalm}, \citenamefont {Zhang}, \citenamefont {Nam}, \citenamefont {Brunner}, \citenamefont {Lim}, \citenamefont {Gisin},\ and\ \citenamefont {Kwiat}}]{Christensen2013}%
  \BibitemOpen
  \bibfield  {author} {\bibinfo {author} {\bibfnamefont {B.~G.}\ \bibnamefont {Christensen}}, \bibinfo {author} {\bibfnamefont {K.~T.}\ \bibnamefont {McCusker}}, \bibinfo {author} {\bibfnamefont {J.~B.}\ \bibnamefont {Altepeter}}, \bibinfo {author} {\bibfnamefont {B.}~\bibnamefont {Calkins}}, \bibinfo {author} {\bibfnamefont {T.}~\bibnamefont {Gerrits}}, \bibinfo {author} {\bibfnamefont {A.~E.}\ \bibnamefont {Lita}}, \bibinfo {author} {\bibfnamefont {A.}~\bibnamefont {Miller}}, \bibinfo {author} {\bibfnamefont {L.~K.}\ \bibnamefont {Shalm}}, \bibinfo {author} {\bibfnamefont {Y.}~\bibnamefont {Zhang}}, \bibinfo {author} {\bibfnamefont {S.~W.}\ \bibnamefont {Nam}}, \bibinfo {author} {\bibfnamefont {N.}~\bibnamefont {Brunner}}, \bibinfo {author} {\bibfnamefont {C.~C.~W.}\ \bibnamefont {Lim}}, \bibinfo {author} {\bibfnamefont {N.}~\bibnamefont {Gisin}}, \ and\ \bibinfo {author} {\bibfnamefont {P.~G.}\ \bibnamefont {Kwiat}},\ }\bibfield  {title} {\enquote {\bibinfo {title} {Detection-loophole-free test of quantum
  nonlocality, and applications},}\ }\href {\doibase 10.1103/PhysRevLett.111.130406} {\bibfield  {journal} {\bibinfo  {journal} {Phys. Rev. Lett.}\ }\textbf {\bibinfo {volume} {111}},\ \bibinfo {pages} {130406} (\bibinfo {year} {2013})}\BibitemShut {NoStop}%
\bibitem [{\citenamefont {Shadbolt}\ \emph {et~al.}(2014)\citenamefont {Shadbolt}, \citenamefont {Mathews}, \citenamefont {Laing},\ and\ \citenamefont {O'Brien}}]{Shadbolt2014}%
  \BibitemOpen
  \bibfield  {author} {\bibinfo {author} {\bibfnamefont {P.}~\bibnamefont {Shadbolt}}, \bibinfo {author} {\bibfnamefont {J.~C.~F.}\ \bibnamefont {Mathews}}, \bibinfo {author} {\bibfnamefont {A.}~\bibnamefont {Laing}}, \ and\ \bibinfo {author} {\bibfnamefont {J.~L.}\ \bibnamefont {O'Brien}},\ }\bibfield  {title} {\enquote {\bibinfo {title} {Testing foundations of quantum mechanics with photons},}\ }\href {\doibase 10.1038/nphys2931} {\bibfield  {journal} {\bibinfo  {journal} {Nature Physics}\ }\textbf {\bibinfo {volume} {10}},\ \bibinfo {pages} {278--286} (\bibinfo {year} {2014})}\BibitemShut {NoStop}%
\bibitem [{\citenamefont {Prevedel}\ \emph {et~al.}(2007)\citenamefont {Prevedel}, \citenamefont {Walther}, \citenamefont {Tiefenbacher}, \citenamefont {B{\"o}hi}, \citenamefont {Kaltenbaek}, \citenamefont {Jennewein},\ and\ \citenamefont {Zeilinger}}]{Prevedel2007}%
  \BibitemOpen
  \bibfield  {author} {\bibinfo {author} {\bibfnamefont {R.}~\bibnamefont {Prevedel}}, \bibinfo {author} {\bibfnamefont {P.}~\bibnamefont {Walther}}, \bibinfo {author} {\bibfnamefont {F.}~\bibnamefont {Tiefenbacher}}, \bibinfo {author} {\bibfnamefont {P.}~\bibnamefont {B{\"o}hi}}, \bibinfo {author} {\bibfnamefont {R.}~\bibnamefont {Kaltenbaek}}, \bibinfo {author} {\bibfnamefont {T.}~\bibnamefont {Jennewein}}, \ and\ \bibinfo {author} {\bibfnamefont {A.}~\bibnamefont {Zeilinger}},\ }\bibfield  {title} {\enquote {\bibinfo {title} {High-speed linear optics quantum computing using active feed-forward},}\ }\href {\doibase 10.1038/nature05346} {\bibfield  {journal} {\bibinfo  {journal} {Nature}\ }\textbf {\bibinfo {volume} {445}},\ \bibinfo {pages} {65--69} (\bibinfo {year} {2007})}\BibitemShut {NoStop}%
\bibitem [{\citenamefont {Kok}(2016)}]{Kok2016}%
  \BibitemOpen
  \bibfield  {author} {\bibinfo {author} {\bibfnamefont {P.}~\bibnamefont {Kok}},\ }\bibfield  {title} {\enquote {\bibinfo {title} {Photonic quantum information processing},}\ }\href {\doibase 10.1080/00107514.2016.1178472} {\bibfield  {journal} {\bibinfo  {journal} {Contemporary Physics}\ }\textbf {\bibinfo {volume} {57}},\ \bibinfo {pages} {526--544} (\bibinfo {year} {2016})}\BibitemShut {NoStop}%
\bibitem [{\citenamefont {Flamini}, \citenamefont {Spagnolo},\ and\ \citenamefont {Sciarrino}(2018)}]{Flamini2019}%
  \BibitemOpen
  \bibfield  {author} {\bibinfo {author} {\bibfnamefont {F.}~\bibnamefont {Flamini}}, \bibinfo {author} {\bibfnamefont {N.}~\bibnamefont {Spagnolo}}, \ and\ \bibinfo {author} {\bibfnamefont {F.}~\bibnamefont {Sciarrino}},\ }\bibfield  {title} {\enquote {\bibinfo {title} {Photonic quantum information processing: a review},}\ }\href {\doibase 10.1088/1361-6633/aad5b2} {\bibfield  {journal} {\bibinfo  {journal} {Reports on Progress in Physics}\ }\textbf {\bibinfo {volume} {82}},\ \bibinfo {pages} {016001} (\bibinfo {year} {2018})}\BibitemShut {NoStop}%
\bibitem [{\citenamefont {Whittaker}\ \emph {et~al.}(2017)\citenamefont {Whittaker}, \citenamefont {Erven}, \citenamefont {Neville}, \citenamefont {Berry}, \citenamefont {O’Brien}, \citenamefont {Cable},\ and\ \citenamefont {Matthews}}]{Whittaker2017}%
  \BibitemOpen
  \bibfield  {author} {\bibinfo {author} {\bibfnamefont {R.}~\bibnamefont {Whittaker}}, \bibinfo {author} {\bibfnamefont {C.}~\bibnamefont {Erven}}, \bibinfo {author} {\bibfnamefont {A.}~\bibnamefont {Neville}}, \bibinfo {author} {\bibfnamefont {M.}~\bibnamefont {Berry}}, \bibinfo {author} {\bibfnamefont {J.~L.}\ \bibnamefont {O’Brien}}, \bibinfo {author} {\bibfnamefont {H.}~\bibnamefont {Cable}}, \ and\ \bibinfo {author} {\bibfnamefont {J.~C.~F.}\ \bibnamefont {Matthews}},\ }\bibfield  {title} {\enquote {\bibinfo {title} {Absorption spectroscopy at the ultimate quantum limit from single-photon states},}\ }\href {\doibase 10.1088/1367-2630/aa5512} {\bibfield  {journal} {\bibinfo  {journal} {New Journal of Physics}\ }\textbf {\bibinfo {volume} {19}},\ \bibinfo {pages} {023013} (\bibinfo {year} {2017})}\BibitemShut {NoStop}%
\bibitem [{\citenamefont {Losero}\ \emph {et~al.}(2018)\citenamefont {Losero}, \citenamefont {Ruo-Berchera}, \citenamefont {Meda}, \citenamefont {Avella},\ and\ \citenamefont {Genovese}}]{Losero2018}%
  \BibitemOpen
  \bibfield  {author} {\bibinfo {author} {\bibfnamefont {E.}~\bibnamefont {Losero}}, \bibinfo {author} {\bibfnamefont {I.}~\bibnamefont {Ruo-Berchera}}, \bibinfo {author} {\bibfnamefont {A.}~\bibnamefont {Meda}}, \bibinfo {author} {\bibfnamefont {A.}~\bibnamefont {Avella}}, \ and\ \bibinfo {author} {\bibfnamefont {M.}~\bibnamefont {Genovese}},\ }\bibfield  {title} {\enquote {\bibinfo {title} {Unbiased estimation of an optical loss at the ultimate quantum limit with twin-beams},}\ }\href {\doibase 10.1038/s41598-018-25501-w} {\bibfield  {journal} {\bibinfo  {journal} {Scientific Reports}\ }\textbf {\bibinfo {volume} {8}},\ \bibinfo {pages} {7431} (\bibinfo {year} {2018})}\BibitemShut {NoStop}%
\bibitem [{\citenamefont {Gol’tsman}\ \emph {et~al.}(2001)\citenamefont {Gol’tsman}, \citenamefont {Okunev}, \citenamefont {Chulkova}, \citenamefont {Lipatov}, \citenamefont {Semenov}, \citenamefont {Smirnov}, \citenamefont {Voronov}, \citenamefont {Dzardanov}, \citenamefont {Williams},\ and\ \citenamefont {Sobolewski}}]{first_snspd}%
  \BibitemOpen
  \bibfield  {author} {\bibinfo {author} {\bibfnamefont {G.~N.}\ \bibnamefont {Gol’tsman}}, \bibinfo {author} {\bibfnamefont {O.}~\bibnamefont {Okunev}}, \bibinfo {author} {\bibfnamefont {G.}~\bibnamefont {Chulkova}}, \bibinfo {author} {\bibfnamefont {A.}~\bibnamefont {Lipatov}}, \bibinfo {author} {\bibfnamefont {A.}~\bibnamefont {Semenov}}, \bibinfo {author} {\bibfnamefont {K.}~\bibnamefont {Smirnov}}, \bibinfo {author} {\bibfnamefont {B.}~\bibnamefont {Voronov}}, \bibinfo {author} {\bibfnamefont {A.}~\bibnamefont {Dzardanov}}, \bibinfo {author} {\bibfnamefont {C.}~\bibnamefont {Williams}}, \ and\ \bibinfo {author} {\bibfnamefont {R.}~\bibnamefont {Sobolewski}},\ }\bibfield  {title} {\enquote {\bibinfo {title} {Picosecond superconducting single-photon optical detector},}\ }\href {\doibase 10.1063/1.1388868} {\bibfield  {journal} {\bibinfo  {journal} {Applied Physics Letters}\ }\textbf {\bibinfo {volume} {79}},\ \bibinfo {pages} {705--707} (\bibinfo {year} {2001})}\BibitemShut {NoStop}%
\bibitem [{\citenamefont {Eisaman}\ \emph {et~al.}(2011)\citenamefont {Eisaman}, \citenamefont {Fan}, \citenamefont {Migdall},\ and\ \citenamefont {Polyakov}}]{Single-photon_source_and_detectors}%
  \BibitemOpen
  \bibfield  {author} {\bibinfo {author} {\bibfnamefont {M.~D.}\ \bibnamefont {Eisaman}}, \bibinfo {author} {\bibfnamefont {J.}~\bibnamefont {Fan}}, \bibinfo {author} {\bibfnamefont {A.}~\bibnamefont {Migdall}}, \ and\ \bibinfo {author} {\bibfnamefont {S.~V.}\ \bibnamefont {Polyakov}},\ }\bibfield  {title} {\enquote {\bibinfo {title} {Invited review article: Single-photon sources and detectors},}\ }\href {\doibase 10.1063/1.3610677} {\bibfield  {journal} {\bibinfo  {journal} {Review of Scientific Instruments}\ }\textbf {\bibinfo {volume} {82}},\ \bibinfo {pages} {071101} (\bibinfo {year} {2011})}\BibitemShut {NoStop}%
\bibitem [{\citenamefont {Hadfield}\ and\ \citenamefont {Johansson}(2016)}]{Superconducting_Devices_in_Quantum_Optics}%
  \BibitemOpen
  \bibfield  {author} {\bibinfo {author} {\bibfnamefont {R.~H.}\ \bibnamefont {Hadfield}}\ and\ \bibinfo {author} {\bibfnamefont {G.}~\bibnamefont {Johansson}},\ }\href {\doibase https://doi.org/10.1007/978-3-319-24091-6} {\emph {\bibinfo {title} {Superconducting Devices in Quantum Optics}}}\ (\bibinfo  {publisher} {Springer Cham},\ \bibinfo {year} {2016})\BibitemShut {NoStop}%
\bibitem [{\citenamefont {Maring}\ \emph {et~al.}(2017)\citenamefont {Maring}, \citenamefont {Farrera}, \citenamefont {Kutluer}, \citenamefont {Mazzera}, \citenamefont {Heinze},\ and\ \citenamefont {de~Riedmatten}}]{Maring2017}%
  \BibitemOpen
  \bibfield  {author} {\bibinfo {author} {\bibfnamefont {N.}~\bibnamefont {Maring}}, \bibinfo {author} {\bibfnamefont {P.}~\bibnamefont {Farrera}}, \bibinfo {author} {\bibfnamefont {K.}~\bibnamefont {Kutluer}}, \bibinfo {author} {\bibfnamefont {M.}~\bibnamefont {Mazzera}}, \bibinfo {author} {\bibfnamefont {G.}~\bibnamefont {Heinze}}, \ and\ \bibinfo {author} {\bibfnamefont {H.}~\bibnamefont {de~Riedmatten}},\ }\bibfield  {title} {\enquote {\bibinfo {title} {Photonic quantum state transfer between a cold atomic gas and a crystal},}\ }\href {\doibase 10.1038/nature24468} {\bibfield  {journal} {\bibinfo  {journal} {Nature}\ }\textbf {\bibinfo {volume} {551}},\ \bibinfo {pages} {485--488} (\bibinfo {year} {2017})}\BibitemShut {NoStop}%
\bibitem [{\citenamefont {Chen}\ \emph {et~al.}(2020)\citenamefont {Chen}, \citenamefont {Zhang}, \citenamefont {Liu}, \citenamefont {Jiang}, \citenamefont {Zhang}, \citenamefont {Hu}, \citenamefont {Guan}, \citenamefont {Yu}, \citenamefont {Xu}, \citenamefont {Lin}, \citenamefont {Li}, \citenamefont {Chen}, \citenamefont {Li}, \citenamefont {You}, \citenamefont {Wang}, \citenamefont {Wang}, \citenamefont {Zhang},\ and\ \citenamefont {Pan}}]{Chen2020}%
  \BibitemOpen
  \bibfield  {author} {\bibinfo {author} {\bibfnamefont {J.-P.}\ \bibnamefont {Chen}}, \bibinfo {author} {\bibfnamefont {C.}~\bibnamefont {Zhang}}, \bibinfo {author} {\bibfnamefont {Y.}~\bibnamefont {Liu}}, \bibinfo {author} {\bibfnamefont {C.}~\bibnamefont {Jiang}}, \bibinfo {author} {\bibfnamefont {W.}~\bibnamefont {Zhang}}, \bibinfo {author} {\bibfnamefont {X.-L.}\ \bibnamefont {Hu}}, \bibinfo {author} {\bibfnamefont {J.-Y.}\ \bibnamefont {Guan}}, \bibinfo {author} {\bibfnamefont {Z.-W.}\ \bibnamefont {Yu}}, \bibinfo {author} {\bibfnamefont {H.}~\bibnamefont {Xu}}, \bibinfo {author} {\bibfnamefont {J.}~\bibnamefont {Lin}}, \bibinfo {author} {\bibfnamefont {M.-J.}\ \bibnamefont {Li}}, \bibinfo {author} {\bibfnamefont {H.}~\bibnamefont {Chen}}, \bibinfo {author} {\bibfnamefont {H.}~\bibnamefont {Li}}, \bibinfo {author} {\bibfnamefont {L.}~\bibnamefont {You}}, \bibinfo {author} {\bibfnamefont {Z.}~\bibnamefont {Wang}}, \bibinfo {author} {\bibfnamefont {X.-B.}\ \bibnamefont {Wang}}, \bibinfo {author}
  {\bibfnamefont {Q.}~\bibnamefont {Zhang}}, \ and\ \bibinfo {author} {\bibfnamefont {J.-W.}\ \bibnamefont {Pan}},\ }\bibfield  {title} {\enquote {\bibinfo {title} {Sending-or-not-sending with independent lasers: Secure twin-field quantum key distribution over 509 km},}\ }\href {\doibase 10.1103/PhysRevLett.124.070501} {\bibfield  {journal} {\bibinfo  {journal} {Phys. Rev. Lett.}\ }\textbf {\bibinfo {volume} {124}},\ \bibinfo {pages} {070501} (\bibinfo {year} {2020})}\BibitemShut {NoStop}%
\bibitem [{\citenamefont {Wei}\ \emph {et~al.}(2022)\citenamefont {Wei}, \citenamefont {Jing}, \citenamefont {Zhang}, \citenamefont {Liao}, \citenamefont {Yuan}, \citenamefont {Fan}, \citenamefont {Lyu}, \citenamefont {Zhou}, \citenamefont {Wang}, \citenamefont {Deng}, \citenamefont {Song}, \citenamefont {Oblak}, \citenamefont {Guo},\ and\ \citenamefont {Zhou}}]{Wei2022}%
  \BibitemOpen
  \bibfield  {author} {\bibinfo {author} {\bibfnamefont {S.-H.}\ \bibnamefont {Wei}}, \bibinfo {author} {\bibfnamefont {B.}~\bibnamefont {Jing}}, \bibinfo {author} {\bibfnamefont {X.-Y.}\ \bibnamefont {Zhang}}, \bibinfo {author} {\bibfnamefont {J.-Y.}\ \bibnamefont {Liao}}, \bibinfo {author} {\bibfnamefont {C.-Z.}\ \bibnamefont {Yuan}}, \bibinfo {author} {\bibfnamefont {B.-Y.}\ \bibnamefont {Fan}}, \bibinfo {author} {\bibfnamefont {C.}~\bibnamefont {Lyu}}, \bibinfo {author} {\bibfnamefont {D.-L.}\ \bibnamefont {Zhou}}, \bibinfo {author} {\bibfnamefont {Y.}~\bibnamefont {Wang}}, \bibinfo {author} {\bibfnamefont {G.-W.}\ \bibnamefont {Deng}}, \bibinfo {author} {\bibfnamefont {H.-Z.}\ \bibnamefont {Song}}, \bibinfo {author} {\bibfnamefont {D.}~\bibnamefont {Oblak}}, \bibinfo {author} {\bibfnamefont {G.-C.}\ \bibnamefont {Guo}}, \ and\ \bibinfo {author} {\bibfnamefont {Q.}~\bibnamefont {Zhou}},\ }\bibfield  {title} {\enquote {\bibinfo {title} {Towards real-world quantum networks: A review},}\ }\href {\doibase
  https://doi.org/10.1002/lpor.202100219} {\bibfield  {journal} {\bibinfo  {journal} {Laser \& Photonics Reviews}\ }\textbf {\bibinfo {volume} {16}},\ \bibinfo {pages} {2100219} (\bibinfo {year} {2022})}\BibitemShut {NoStop}%
\bibitem [{\citenamefont {van Leent}\ \emph {et~al.}(2022)\citenamefont {van Leent}, \citenamefont {Bock}, \citenamefont {Fertig}, \citenamefont {Garthoff}, \citenamefont {Eppelt}, \citenamefont {Zhou}, \citenamefont {Malik}, \citenamefont {Seubert}, \citenamefont {Bauer}, \citenamefont {Rosenfeld}, \citenamefont {Zhang}, \citenamefont {Becher},\ and\ \citenamefont {Weinfurter}}]{vanLeent2022}%
  \BibitemOpen
  \bibfield  {author} {\bibinfo {author} {\bibfnamefont {T.}~\bibnamefont {van Leent}}, \bibinfo {author} {\bibfnamefont {M.}~\bibnamefont {Bock}}, \bibinfo {author} {\bibfnamefont {F.}~\bibnamefont {Fertig}}, \bibinfo {author} {\bibfnamefont {R.}~\bibnamefont {Garthoff}}, \bibinfo {author} {\bibfnamefont {S.}~\bibnamefont {Eppelt}}, \bibinfo {author} {\bibfnamefont {Y.}~\bibnamefont {Zhou}}, \bibinfo {author} {\bibfnamefont {P.}~\bibnamefont {Malik}}, \bibinfo {author} {\bibfnamefont {M.}~\bibnamefont {Seubert}}, \bibinfo {author} {\bibfnamefont {T.}~\bibnamefont {Bauer}}, \bibinfo {author} {\bibfnamefont {W.}~\bibnamefont {Rosenfeld}}, \bibinfo {author} {\bibfnamefont {W.}~\bibnamefont {Zhang}}, \bibinfo {author} {\bibfnamefont {C.}~\bibnamefont {Becher}}, \ and\ \bibinfo {author} {\bibfnamefont {H.}~\bibnamefont {Weinfurter}},\ }\bibfield  {title} {\enquote {\bibinfo {title} {Entangling single atoms over 33 km telecom fibre},}\ }\href {\doibase 10.1038/s41586-022-04764-4} {\bibfield  {journal} {\bibinfo
  {journal} {Nature}\ }\textbf {\bibinfo {volume} {607}},\ \bibinfo {pages} {69--73} (\bibinfo {year} {2022})}\BibitemShut {NoStop}%
\bibitem [{\citenamefont {James}\ \emph {et~al.}(2001)\citenamefont {James}, \citenamefont {Kwiat}, \citenamefont {Munro},\ and\ \citenamefont {White}}]{QST2001}%
  \BibitemOpen
  \bibfield  {author} {\bibinfo {author} {\bibfnamefont {D.~F.~V.}\ \bibnamefont {James}}, \bibinfo {author} {\bibfnamefont {P.~G.}\ \bibnamefont {Kwiat}}, \bibinfo {author} {\bibfnamefont {W.~J.}\ \bibnamefont {Munro}}, \ and\ \bibinfo {author} {\bibfnamefont {A.~G.}\ \bibnamefont {White}},\ }\bibfield  {title} {\enquote {\bibinfo {title} {Measurement of qubits},}\ }\href {\doibase 10.1103/PhysRevA.64.052312} {\bibfield  {journal} {\bibinfo  {journal} {Phys. Rev. A}\ }\textbf {\bibinfo {volume} {64}},\ \bibinfo {pages} {052312} (\bibinfo {year} {2001})}\BibitemShut {NoStop}%
\bibitem [{\citenamefont {Altepeter}, \citenamefont {Jeffrey},\ and\ \citenamefont {Kwiat}(2005)}]{Altepeter2005}%
  \BibitemOpen
  \bibfield  {author} {\bibinfo {author} {\bibfnamefont {J.}~\bibnamefont {Altepeter}}, \bibinfo {author} {\bibfnamefont {E.}~\bibnamefont {Jeffrey}}, \ and\ \bibinfo {author} {\bibfnamefont {P.}~\bibnamefont {Kwiat}},\ }\bibfield  {title} {\enquote {\bibinfo {title} {Photonic state tomography},}\ \ }(\bibinfo  {publisher} {Academic Press},\ \bibinfo {year} {2005})\ pp.\ \bibinfo {pages} {105--159}\BibitemShut {NoStop}%
\bibitem [{\citenamefont {Söllner}\ \emph {et~al.}(2011)\citenamefont {Söllner}, \citenamefont {Gschösser}, \citenamefont {Mai}, \citenamefont {Pressl}, \citenamefont {Vörös},\ and\ \citenamefont {Weihs}}]{born2011}%
  \BibitemOpen
  \bibfield  {author} {\bibinfo {author} {\bibfnamefont {I.}~\bibnamefont {Söllner}}, \bibinfo {author} {\bibfnamefont {B.}~\bibnamefont {Gschösser}}, \bibinfo {author} {\bibfnamefont {P.}~\bibnamefont {Mai}}, \bibinfo {author} {\bibfnamefont {B.}~\bibnamefont {Pressl}}, \bibinfo {author} {\bibfnamefont {Z.}~\bibnamefont {Vörös}}, \ and\ \bibinfo {author} {\bibfnamefont {G.}~\bibnamefont {Weihs}},\ }\bibfield  {title} {\enquote {\bibinfo {title} {Testing born's rule in quantum mechanics for three mutually exclusive events},}\ }\href {\doibase 10.1007/s10701-011-9597-5} {\bibfield  {journal} {\bibinfo  {journal} {Foundations of Physics}\ }\textbf {\bibinfo {volume} {42}},\ \bibinfo {pages} {742--751} (\bibinfo {year} {2011})}\BibitemShut {NoStop}%
\bibitem [{\citenamefont {Kauten}\ \emph {et~al.}(2017)\citenamefont {Kauten}, \citenamefont {Keil}, \citenamefont {Kaufmann}, \citenamefont {Pressl}, \citenamefont {Brukner},\ and\ \citenamefont {Weihs}}]{Kauten_2017_5Path}%
  \BibitemOpen
  \bibfield  {author} {\bibinfo {author} {\bibfnamefont {T.}~\bibnamefont {Kauten}}, \bibinfo {author} {\bibfnamefont {R.}~\bibnamefont {Keil}}, \bibinfo {author} {\bibfnamefont {T.}~\bibnamefont {Kaufmann}}, \bibinfo {author} {\bibfnamefont {B.}~\bibnamefont {Pressl}}, \bibinfo {author} {\bibfnamefont {{\v{C} }.}~\bibnamefont {Brukner}}, \ and\ \bibinfo {author} {\bibfnamefont {G.}~\bibnamefont {Weihs}},\ }\bibfield  {title} {\enquote {\bibinfo {title} {Obtaining tight bounds on higher-order interferences with a 5-path interferometer},}\ }\href {\doibase 10.1088/1367-2630/aa5d98} {\bibfield  {journal} {\bibinfo  {journal} {New Journal of Physics}\ }\textbf {\bibinfo {volume} {19}},\ \bibinfo {pages} {033017} (\bibinfo {year} {2017})}\BibitemShut {NoStop}%
\bibitem [{\citenamefont {Rozema}\ \emph {et~al.}(2021)\citenamefont {Rozema}, \citenamefont {Zhuo}, \citenamefont {Paterek},\ and\ \citenamefont {Daki\ifmmode~\acute{c}\else \'{c}\fi{}}}]{Rozema2021}%
  \BibitemOpen
  \bibfield  {author} {\bibinfo {author} {\bibfnamefont {L.~A.}\ \bibnamefont {Rozema}}, \bibinfo {author} {\bibfnamefont {Z.}~\bibnamefont {Zhuo}}, \bibinfo {author} {\bibfnamefont {T.}~\bibnamefont {Paterek}}, \ and\ \bibinfo {author} {\bibfnamefont {B.}~\bibnamefont {Daki\ifmmode~\acute{c}\else \'{c}\fi{}}},\ }\bibfield  {title} {\enquote {\bibinfo {title} {Higher-order interference between multiple quantum particles interacting nonlinearly},}\ }\href {\doibase 10.1103/PhysRevA.103.052204} {\bibfield  {journal} {\bibinfo  {journal} {Phys. Rev. A}\ }\textbf {\bibinfo {volume} {103}},\ \bibinfo {pages} {052204} (\bibinfo {year} {2021})}\BibitemShut {NoStop}%
\bibitem [{\citenamefont {Vogl}\ \emph {et~al.}(2021)\citenamefont {Vogl}, \citenamefont {Knopf}, \citenamefont {Weissflog}, \citenamefont {Lam},\ and\ \citenamefont {Eilenberger}}]{Vogl2021}%
  \BibitemOpen
  \bibfield  {author} {\bibinfo {author} {\bibfnamefont {T.}~\bibnamefont {Vogl}}, \bibinfo {author} {\bibfnamefont {H.}~\bibnamefont {Knopf}}, \bibinfo {author} {\bibfnamefont {M.}~\bibnamefont {Weissflog}}, \bibinfo {author} {\bibfnamefont {P.~K.}\ \bibnamefont {Lam}}, \ and\ \bibinfo {author} {\bibfnamefont {F.}~\bibnamefont {Eilenberger}},\ }\bibfield  {title} {\enquote {\bibinfo {title} {Sensitive single-photon test of extended quantum theory with two-dimensional hexagonal boron nitride},}\ }\href {\doibase 10.1103/PhysRevResearch.3.013296} {\bibfield  {journal} {\bibinfo  {journal} {Phys. Rev. Res.}\ }\textbf {\bibinfo {volume} {3}},\ \bibinfo {pages} {013296} (\bibinfo {year} {2021})}\BibitemShut {NoStop}%
\bibitem [{\citenamefont {Gstir}\ \emph {et~al.}(2021)\citenamefont {Gstir}, \citenamefont {Chan}, \citenamefont {Eichelkraut}, \citenamefont {Szameit}, \citenamefont {Keil},\ and\ \citenamefont {Weihs}}]{Gstir_2021}%
  \BibitemOpen
  \bibfield  {author} {\bibinfo {author} {\bibfnamefont {S.}~\bibnamefont {Gstir}}, \bibinfo {author} {\bibfnamefont {E.}~\bibnamefont {Chan}}, \bibinfo {author} {\bibfnamefont {T.}~\bibnamefont {Eichelkraut}}, \bibinfo {author} {\bibfnamefont {A.}~\bibnamefont {Szameit}}, \bibinfo {author} {\bibfnamefont {R.}~\bibnamefont {Keil}}, \ and\ \bibinfo {author} {\bibfnamefont {G.}~\bibnamefont {Weihs}},\ }\bibfield  {title} {\enquote {\bibinfo {title} {Towards probing for hypercomplex quantum mechanics in a waveguide interferometer},}\ }\href {\doibase 10.1088/1367-2630/ac2451} {\bibfield  {journal} {\bibinfo  {journal} {New Journal of Physics}\ }\textbf {\bibinfo {volume} {23}},\ \bibinfo {pages} {093038} (\bibinfo {year} {2021})}\BibitemShut {NoStop}%
\bibitem [{\citenamefont {Gstir}(2023)}]{Gstir2023}%
  \BibitemOpen
  \bibfield  {author} {\bibinfo {author} {\bibfnamefont {S.}~\bibnamefont {Gstir}},\ }\emph {\bibinfo {title} {Waveguide Interferometers for Fundamental Investigation of Quantum Mechanics}},\ \href {https://resolver.obvsg.at/urn:nbn:at:at-ubi:1-128187} {Ph.D. thesis},\ \bibinfo  {school} {University of Innsbruck} (\bibinfo {year} {2023})\BibitemShut {NoStop}%
\bibitem [{\citenamefont {Kerman}\ \emph {et~al.}(2006)\citenamefont {Kerman}, \citenamefont {Dauler}, \citenamefont {Keicher}, \citenamefont {Yang}, \citenamefont {Berggren}, \citenamefont {Gol’tsman},\ and\ \citenamefont {Voronov}}]{Kerman2006}%
  \BibitemOpen
  \bibfield  {author} {\bibinfo {author} {\bibfnamefont {A.~J.}\ \bibnamefont {Kerman}}, \bibinfo {author} {\bibfnamefont {E.~A.}\ \bibnamefont {Dauler}}, \bibinfo {author} {\bibfnamefont {W.~E.}\ \bibnamefont {Keicher}}, \bibinfo {author} {\bibfnamefont {J.~K.~W.}\ \bibnamefont {Yang}}, \bibinfo {author} {\bibfnamefont {K.~K.}\ \bibnamefont {Berggren}}, \bibinfo {author} {\bibfnamefont {G.}~\bibnamefont {Gol’tsman}}, \ and\ \bibinfo {author} {\bibfnamefont {B.}~\bibnamefont {Voronov}},\ }\bibfield  {title} {\enquote {\bibinfo {title} {{Kinetic-inductance-limited reset time of superconducting nanowire photon counters}},}\ }\href {\doibase 10.1063/1.2183810} {\bibfield  {journal} {\bibinfo  {journal} {Applied Physics Letters}\ }\textbf {\bibinfo {volume} {88}},\ \bibinfo {pages} {111116} (\bibinfo {year} {2006})}\BibitemShut {NoStop}%
\bibitem [{\citenamefont {Ware}\ \emph {et~al.}(2007)\citenamefont {Ware}, \citenamefont {Migdall}, \citenamefont {Bienfang},\ and\ \citenamefont {Polyakov}}]{Ware2007}%
  \BibitemOpen
  \bibfield  {author} {\bibinfo {author} {\bibfnamefont {M.}~\bibnamefont {Ware}}, \bibinfo {author} {\bibfnamefont {A.}~\bibnamefont {Migdall}}, \bibinfo {author} {\bibfnamefont {J.~C.}\ \bibnamefont {Bienfang}}, \ and\ \bibinfo {author} {\bibfnamefont {S.~V.}\ \bibnamefont {Polyakov}},\ }\bibfield  {title} {\enquote {\bibinfo {title} {Calibrating photon-counting detectors to high accuracy: background and deadtime issues},}\ }\href {\doibase 10.1080/09500340600759597} {\bibfield  {journal} {\bibinfo  {journal} {Journal of Modern Optics}\ }\textbf {\bibinfo {volume} {54}},\ \bibinfo {pages} {361--372} (\bibinfo {year} {2007})}\BibitemShut {NoStop}%
\bibitem [{\citenamefont {Kornilov}(2014)}]{nonlin_pd_Kornilov:14}%
  \BibitemOpen
  \bibfield  {author} {\bibinfo {author} {\bibfnamefont {V.}~\bibnamefont {Kornilov}},\ }\bibfield  {title} {\enquote {\bibinfo {title} {Effects of dead time and afterpulses in photon detector on measured statistics of stochastic radiation},}\ }\href {\doibase 10.1364/JOSAA.31.000007} {\bibfield  {journal} {\bibinfo  {journal} {J. Opt. Soc. Am. A}\ }\textbf {\bibinfo {volume} {31}},\ \bibinfo {pages} {7--15} (\bibinfo {year} {2014})}\BibitemShut {NoStop}%
\bibitem [{\citenamefont {Kauten}\ \emph {et~al.}(2014)\citenamefont {Kauten}, \citenamefont {Pressl}, \citenamefont {Kaufmann},\ and\ \citenamefont {Weihs}}]{kautenNonlin}%
  \BibitemOpen
  \bibfield  {author} {\bibinfo {author} {\bibfnamefont {T.}~\bibnamefont {Kauten}}, \bibinfo {author} {\bibfnamefont {B.}~\bibnamefont {Pressl}}, \bibinfo {author} {\bibfnamefont {T.}~\bibnamefont {Kaufmann}}, \ and\ \bibinfo {author} {\bibfnamefont {G.}~\bibnamefont {Weihs}},\ }\bibfield  {title} {\enquote {\bibinfo {title} {Measurement and modeling of the nonlinearity of photovoltaic and geiger-mode photodiodes},}\ }\href {\doibase 10.1063/1.4879820} {\bibfield  {journal} {\bibinfo  {journal} {Review of Scientific Instruments}\ }\textbf {\bibinfo {volume} {85}},\ \bibinfo {pages} {063102} (\bibinfo {year} {2014})}\BibitemShut {NoStop}%
\bibitem [{\citenamefont {Hloušek}, \citenamefont {Straka},\ and\ \citenamefont {Ježek}(2023)}]{Hlousek2023}%
  \BibitemOpen
  \bibfield  {author} {\bibinfo {author} {\bibfnamefont {J.}~\bibnamefont {Hloušek}}, \bibinfo {author} {\bibfnamefont {I.}~\bibnamefont {Straka}}, \ and\ \bibinfo {author} {\bibfnamefont {M.}~\bibnamefont {Ježek}},\ }\bibfield  {title} {\enquote {\bibinfo {title} {{Experimental observation of anomalous supralinear response of single-photon detectors}},}\ }\href {\doibase 10.1063/5.0106987} {\bibfield  {journal} {\bibinfo  {journal} {Applied Physics Reviews}\ }\textbf {\bibinfo {volume} {10}},\ \bibinfo {pages} {011412} (\bibinfo {year} {2023})}\BibitemShut {NoStop}%
\bibitem [{\citenamefont {Zhou}\ \emph {et~al.}(2013)\citenamefont {Zhou}, \citenamefont {Frucci}, \citenamefont {Mattioli}, \citenamefont {Gaggero}, \citenamefont {Leoni}, \citenamefont {Jahanmirinejad}, \citenamefont {Hoang},\ and\ \citenamefont {Fiore}}]{N_Photon_autocorrelator}%
  \BibitemOpen
  \bibfield  {author} {\bibinfo {author} {\bibfnamefont {Z.}~\bibnamefont {Zhou}}, \bibinfo {author} {\bibfnamefont {G.}~\bibnamefont {Frucci}}, \bibinfo {author} {\bibfnamefont {F.}~\bibnamefont {Mattioli}}, \bibinfo {author} {\bibfnamefont {A.}~\bibnamefont {Gaggero}}, \bibinfo {author} {\bibfnamefont {R.}~\bibnamefont {Leoni}}, \bibinfo {author} {\bibfnamefont {S.}~\bibnamefont {Jahanmirinejad}}, \bibinfo {author} {\bibfnamefont {T.~B.}\ \bibnamefont {Hoang}}, \ and\ \bibinfo {author} {\bibfnamefont {A.}~\bibnamefont {Fiore}},\ }\bibfield  {title} {\enquote {\bibinfo {title} {Ultrasensitive n-photon interferometric autocorrelator},}\ }\href {\doibase 10.1103/PhysRevLett.110.133605} {\bibfield  {journal} {\bibinfo  {journal} {Phys. Rev. Lett.}\ }\textbf {\bibinfo {volume} {110}},\ \bibinfo {pages} {133605} (\bibinfo {year} {2013})}\BibitemShut {NoStop}%
\bibitem [{\citenamefont {Marsili}\ \emph {et~al.}(2016)\citenamefont {Marsili}, \citenamefont {Stevens}, \citenamefont {Kozorezov}, \citenamefont {Verma}, \citenamefont {Lambert}, \citenamefont {Stern}, \citenamefont {Horansky}, \citenamefont {Dyer}, \citenamefont {Duff}, \citenamefont {Pappas}, \citenamefont {Lita}, \citenamefont {Shaw}, \citenamefont {Mirin},\ and\ \citenamefont {Nam}}]{P_click}%
  \BibitemOpen
  \bibfield  {author} {\bibinfo {author} {\bibfnamefont {F.}~\bibnamefont {Marsili}}, \bibinfo {author} {\bibfnamefont {M.~J.}\ \bibnamefont {Stevens}}, \bibinfo {author} {\bibfnamefont {A.}~\bibnamefont {Kozorezov}}, \bibinfo {author} {\bibfnamefont {V.~B.}\ \bibnamefont {Verma}}, \bibinfo {author} {\bibfnamefont {C.}~\bibnamefont {Lambert}}, \bibinfo {author} {\bibfnamefont {J.~A.}\ \bibnamefont {Stern}}, \bibinfo {author} {\bibfnamefont {R.~D.}\ \bibnamefont {Horansky}}, \bibinfo {author} {\bibfnamefont {S.}~\bibnamefont {Dyer}}, \bibinfo {author} {\bibfnamefont {S.}~\bibnamefont {Duff}}, \bibinfo {author} {\bibfnamefont {D.~P.}\ \bibnamefont {Pappas}}, \bibinfo {author} {\bibfnamefont {A.~E.}\ \bibnamefont {Lita}}, \bibinfo {author} {\bibfnamefont {M.~D.}\ \bibnamefont {Shaw}}, \bibinfo {author} {\bibfnamefont {R.~P.}\ \bibnamefont {Mirin}}, \ and\ \bibinfo {author} {\bibfnamefont {S.~W.}\ \bibnamefont {Nam}},\ }\bibfield  {title} {\enquote {\bibinfo {title} {Hotspot relaxation dynamics in a
  current-carrying superconductor},}\ }\href {\doibase 10.1103/PhysRevB.93.094518} {\bibfield  {journal} {\bibinfo  {journal} {Phys. Rev. B}\ }\textbf {\bibinfo {volume} {93}},\ \bibinfo {pages} {094518} (\bibinfo {year} {2016})}\BibitemShut {NoStop}%
\bibitem [{\citenamefont {Coslovi}\ and\ \citenamefont {Righini}(1980)}]{fast_nonlin_det_poly_Coslovi:80}%
  \BibitemOpen
  \bibfield  {author} {\bibinfo {author} {\bibfnamefont {L.}~\bibnamefont {Coslovi}}\ and\ \bibinfo {author} {\bibfnamefont {F.}~\bibnamefont {Righini}},\ }\bibfield  {title} {\enquote {\bibinfo {title} {Fast determination of the nonlinearity of photodetectors},}\ }\href {\doibase 10.1364/AO.19.003200} {\bibfield  {journal} {\bibinfo  {journal} {Appl. Opt.}\ }\textbf {\bibinfo {volume} {19}},\ \bibinfo {pages} {3200--3203} (\bibinfo {year} {1980})}\BibitemShut {NoStop}%
\bibitem [{\citenamefont {Autebert}\ \emph {et~al.}(2020)\citenamefont {Autebert}, \citenamefont {Gras}, \citenamefont {Amri}, \citenamefont {Perrenoud}, \citenamefont {Caloz}, \citenamefont {Zbinden},\ and\ \citenamefont {Bussi{\`{e}}res}}]{DirectMeasOfRecoveryTime}%
  \BibitemOpen
  \bibfield  {author} {\bibinfo {author} {\bibfnamefont {C.}~\bibnamefont {Autebert}}, \bibinfo {author} {\bibfnamefont {G.}~\bibnamefont {Gras}}, \bibinfo {author} {\bibfnamefont {E.}~\bibnamefont {Amri}}, \bibinfo {author} {\bibfnamefont {M.}~\bibnamefont {Perrenoud}}, \bibinfo {author} {\bibfnamefont {M.}~\bibnamefont {Caloz}}, \bibinfo {author} {\bibfnamefont {H.}~\bibnamefont {Zbinden}}, \ and\ \bibinfo {author} {\bibfnamefont {F.}~\bibnamefont {Bussi{\`{e}}res}},\ }\bibfield  {title} {\enquote {\bibinfo {title} {Direct measurement of the recovery time of superconducting nanowire single-photon detectors},}\ }\href {\doibase 10.1063/5.0007976} {\bibfield  {journal} {\bibinfo  {journal} {Journal of Applied Physics}\ }\textbf {\bibinfo {volume} {128}},\ \bibinfo {pages} {074504} (\bibinfo {year} {2020})}\BibitemShut {NoStop}%
\bibitem [{\citenamefont {Kerman}\ \emph {et~al.}(2013)\citenamefont {Kerman}, \citenamefont {Rosenberg}, \citenamefont {Molnar},\ and\ \citenamefont {Dauler}}]{AC_biasing}%
  \BibitemOpen
  \bibfield  {author} {\bibinfo {author} {\bibfnamefont {A.~J.}\ \bibnamefont {Kerman}}, \bibinfo {author} {\bibfnamefont {D.}~\bibnamefont {Rosenberg}}, \bibinfo {author} {\bibfnamefont {R.~J.}\ \bibnamefont {Molnar}}, \ and\ \bibinfo {author} {\bibfnamefont {E.~A.}\ \bibnamefont {Dauler}},\ }\bibfield  {title} {\enquote {\bibinfo {title} {Readout of superconducting nanowire single-photon detectors at high count rates},}\ }\href {\doibase 10.1063/1.4799397} {\bibfield  {journal} {\bibinfo  {journal} {Journal of Applied Physics}\ }\textbf {\bibinfo {volume} {113}},\ \bibinfo {pages} {144511} (\bibinfo {year} {2013})}\BibitemShut {NoStop}%
\bibitem [{\citenamefont {Ferrari}\ \emph {et~al.}(2019)\citenamefont {Ferrari}, \citenamefont {Kovalyuk}, \citenamefont {Vetter}, \citenamefont {Lee}, \citenamefont {Rockstuhl}, \citenamefont {Semenov}, \citenamefont {Gol'tsman},\ and\ \citenamefont {Pernice}}]{Ferrari:19_detect_response_AC_biasing}%
  \BibitemOpen
  \bibfield  {author} {\bibinfo {author} {\bibfnamefont {S.}~\bibnamefont {Ferrari}}, \bibinfo {author} {\bibfnamefont {V.}~\bibnamefont {Kovalyuk}}, \bibinfo {author} {\bibfnamefont {A.}~\bibnamefont {Vetter}}, \bibinfo {author} {\bibfnamefont {C.}~\bibnamefont {Lee}}, \bibinfo {author} {\bibfnamefont {C.}~\bibnamefont {Rockstuhl}}, \bibinfo {author} {\bibfnamefont {A.}~\bibnamefont {Semenov}}, \bibinfo {author} {\bibfnamefont {G.}~\bibnamefont {Gol'tsman}}, \ and\ \bibinfo {author} {\bibfnamefont {W.}~\bibnamefont {Pernice}},\ }\bibfield  {title} {\enquote {\bibinfo {title} {Analysis of the detection response of waveguide-integrated superconducting nanowire single-photon detectors at high count rate},}\ }\href {\doibase 10.1063/1.5113652} {\bibfield  {journal} {\bibinfo  {journal} {Applied Physics Letters}\ }\textbf {\bibinfo {volume} {115}},\ \bibinfo {pages} {101104} (\bibinfo {year} {2019})}\BibitemShut {NoStop}%
\bibitem [{Note1()}]{Note1}%
  \BibitemOpen
  \bibinfo {note} {Note that the sign of the residuum $r$ in Eq.~\ref {eq_res_r} is here chosen opposite to Ref.~\protect \rev@citealpnum {kautenNonlin}.}\BibitemShut {Stop}%
\bibitem [{Note2()}]{Note2}%
  \BibitemOpen
  \bibinfo {note} {A $5^{(\protect \text {th})}$-order polynomial interpolation of the data was used to extract $\Delta $ at exactly the same countrate for all bias currents.}\BibitemShut {Stop}%
\bibitem [{Note3()}]{Note3}%
  \BibitemOpen
  \bibinfo {note} {A noise spike in the trailing edge of the pulse can trigger the counting electronics a second time}\BibitemShut {NoStop}%
\bibitem [{\citenamefont {Ferrari}\ \emph {et~al.}(2017)\citenamefont {Ferrari}, \citenamefont {Kovalyuk}, \citenamefont {Hartmann}, \citenamefont {Vetter}, \citenamefont {Kahl}, \citenamefont {Lee}, \citenamefont {Korneev}, \citenamefont {Rockstuhl}, \citenamefont {Gol'tsman},\ and\ \citenamefont {Pernice}}]{Ferrari:17_hs_relax_time}%
  \BibitemOpen
  \bibfield  {author} {\bibinfo {author} {\bibfnamefont {S.}~\bibnamefont {Ferrari}}, \bibinfo {author} {\bibfnamefont {V.}~\bibnamefont {Kovalyuk}}, \bibinfo {author} {\bibfnamefont {W.}~\bibnamefont {Hartmann}}, \bibinfo {author} {\bibfnamefont {A.}~\bibnamefont {Vetter}}, \bibinfo {author} {\bibfnamefont {O.}~\bibnamefont {Kahl}}, \bibinfo {author} {\bibfnamefont {C.}~\bibnamefont {Lee}}, \bibinfo {author} {\bibfnamefont {A.}~\bibnamefont {Korneev}}, \bibinfo {author} {\bibfnamefont {C.}~\bibnamefont {Rockstuhl}}, \bibinfo {author} {\bibfnamefont {G.}~\bibnamefont {Gol'tsman}}, \ and\ \bibinfo {author} {\bibfnamefont {W.}~\bibnamefont {Pernice}},\ }\bibfield  {title} {\enquote {\bibinfo {title} {Hot-spot relaxation time current dependence in niobium nitride waveguide-integrated superconducting nanowire single-photon detectors},}\ }\href {\doibase 10.1364/OE.25.008739} {\bibfield  {journal} {\bibinfo  {journal} {Opt. Express}\ }\textbf {\bibinfo {volume} {25}},\ \bibinfo {pages} {8739--8750} (\bibinfo
  {year} {2017})}\BibitemShut {NoStop}%
\bibitem [{\citenamefont {Blazek}\ and\ \citenamefont {Els\"a\ss{}er}(2011)}]{SLED_g2}%
  \BibitemOpen
  \bibfield  {author} {\bibinfo {author} {\bibfnamefont {M.}~\bibnamefont {Blazek}}\ and\ \bibinfo {author} {\bibfnamefont {W.}~\bibnamefont {Els\"a\ss{}er}},\ }\bibfield  {title} {\enquote {\bibinfo {title} {Coherent and thermal light: Tunable hybrid states with second-order coherence without first-order coherence},}\ }\href {\doibase 10.1103/PhysRevA.84.063840} {\bibfield  {journal} {\bibinfo  {journal} {Phys. Rev. A}\ }\textbf {\bibinfo {volume} {84}},\ \bibinfo {pages} {063840} (\bibinfo {year} {2011})}\BibitemShut {NoStop}%
\end{thebibliography}%

\end{document}